\documentclass{aastex63}
\usepackage{graphicx}
\usepackage{subfigure}
\usepackage{booktabs}
\usepackage{lineno}
\usepackage{longtable}

\shorttitle{huang article}
\shortauthors{huang et al.}

\begin{document}
\setlength{\parskip}{2ex} 

\title{ Comparative Analyses of Plasma Properties and Composition in Two Types of Small-Scale Interplanetary Flux-ropes}


\author[0000-0003-0099-0406]{Jin Huang}
\affiliation{Yunnan Observatories, Chinese Academy of Sciences, Kunming 650011, People's Republic of China}
\affiliation{Shandong Provincial Key Laboratory of Optical Astronomy and Solar-Terrestrial Environment, Shandong University, Weihai 264209, People¡¯s Republic of China}
\affiliation{University of Chinese Academy of Sciences, Beijing 100049, People's Republic of China}

\author{Yu Liu}
\affiliation{Yunnan Observatories, Chinese Academy of Sciences, Kunming 650011, People's Republic of China}
\affiliation{Shandong Provincial Key Laboratory of Optical Astronomy and Solar-Terrestrial Environment, Shandong University, Weihai 264209, People¡¯s Republic of China}
\affiliation{University of Chinese Academy of Sciences, Beijing 100049, People's Republic of China}

\author{Jihong Liu}

\affiliation{College of Physics and Electric Information Engineering , Shijiazhuang University, Shijiazhuang 050035, People's Republic of China}

\author{Yuandeng Shen}
\affiliation{Yunnan Observatories, Chinese Academy of Sciences, Kunming 650011, People's Republic of China}

\nocollaboration{4}

\begin{abstract}

The origin of small-scale interplanetary magnetic flux-ropes (SIMFRs) and the relationship between SIMFRs and magnetic clouds (MCs) are still controversial. In this study, two populations of SMIFRs were collected, i.e., SIMFRs originating from the Sun (SIMFR-SUN) and those originating from the solar wind (SIMFR-SW). We defined the SIMFR-SUN (SIMFR-SW) as the SMIFRs that include (exclude) the counter-streaming suprathermal electrons and stay away from (close to) the heliospheric current sheet. After fitting with force-free flux-rope model, 52 SIMFR-SUN and 57 SIMFR-SW events observed by Advanced Composition Explorer (ACE) from 1998 February to 2011 August were qualified. Using the approach of relating the measurements to their spatial position within the flux-ropes, a comparative survey of plasma and composition characteristics inside the two populations of SIMFRs is presented. Results show that the two populations of SIMFRs have apparent differences. Compared with SIMFR-SW, SIMFR-SUN are MC-like, featuring lower central proton density, higher $V_{rad}$, higher low-FIP element abundances, higher and more fluctuate average ion charge-states and the ion charge-state ratios which are related to the heating in low corona. In addition, for the ion charge-state distributions inside SIMFR-SUN, the sunward side is higher than earthward, which might be caused by the flare heating during eruption. Moreover, both SIMFR-SUN and MCs show anti-correlation between plasma $\beta$ and He/P trend. These characteristics indicate that SIMFR-SUN and MCs are very likely to have the identical origination. This study supports the two-source origin of SIMFRs, i.e., the solar corona and the solar wind.

\end{abstract}


\section{Introduction} \label{sec:intro}

Coronal mass ejections (CMEs) are the most severe explosive phenomena in the heliosphere. The CMEs' interplanetary counterparts are termed as the interplanetary coronal mass ejections (ICMEs). There is a subset of ICMEs that can be fitted with the Lundquist magnetic flux-rope model \citep{1950Ark.35...361}, known as the magnetic clouds (MCs), which are thought to be significantly geoeffective with a critical role in solar-terrestrial effects. MCs are large scale interplanetary magnetic flux-ropes, which have been extensively studied in the literature \citep[e.g.,][]{2003JGRA..108.1370W}. For in situ observations, an MC has been defined as a transient structure which possess enhanced magnetic field strength, large and smooth rotation of the field vector in view of spacecraft passage, low proton temperature ($T_{p}$), low plasma beta ($\beta$), as well as the criteria of ionization levels and composition \citep{2010SoPh..264..189R}. MCs have durations of about a day and diameters of 0.2 au to 0.4 au at the Earth's orbit \citep[e.g.,][]{1990JGR....9511957L}. With the help of white light images of Solar Terrestrial Relations Observatory (STEREO), MCs have been associated with CMEs by continual tracking from the Sun to 1 AU and in situ measurements (e.g., \cite{2011ApJ...734....7R}).

Aside from MCs, there also exist flux-ropes with smaller scale in interplanetary. Compared with MCs, small interplanetary magnetic flux-ropes (SIMFRs) are characterized by short durations (less than 12 hours), small diameters (no more than 0.2 AU), lower magnetic field magnitude $\left|B\right|$, higher $T_{p}$, and larger plasma $\beta$ \citep[e.g.,][]{2015JGRA..12010175F, 2007JGRA..112.2102F,2014JGRA..119.7088J}. They were first identified with the in-situ measurements at 1 au \citep{2000GeoRL..27...57M} two decades ago, much later than the discovery of MCs in 1981 \citep{1981JGR....86.6673B}. Most of SIMFRs are effective for substorms \citep{2010JGRA..115.9108F,2013JASTP..95....1Z}.

However, owing to the weak density fluctuations of small-scale ejecta and considerable distance that separates the lower corona and the spacecraft, the available instruments are difficult to trace back SIMFRs' propagation from the Sun by images. To date, the source of SIMFRs remains controversial, and the relationship between SIMFRs and MCs are still unclear.

On the one hand, \citet{2007JGRA..112.2102F} found that flux-ropes as a whole have a continuous size distribution and thus proposed that SIMFRs possibly are the interplanetary manifestations of small-scale CMEs. In other words, SIMFRs and MCs may have the same source but different scales. As a basis of this view, \citet{2000JGR...10525133W} reported small-scale phenomena in white-light coronal, including outward plasma blobs ejected continually from the cusp-like bases of a coronal streamer. \citet[]{2009SoPh..258..129S} suggested such streamer plasma blobs have the helical structure of MFRs. SIMFRs could also be formed by the erosion of MCs \citep[e.g.,][]{2009ApJ...705.1385F,2015JGRA..120...43R}. There is an observation case that small interplanetary transient traces back to a large CME event \citep{2011ApJ...734....7R}. Recently, \citet{2018A&A...616A..41W} and \citet{2019ApJ...876...57W} indicated respectively that cool prominence material signatures can be found within MCs and SIMFRs. In addition, small-scale ¡°minifilament¡± flux-ropes form when photospheric magnetic flux cancels. \citet{2020ApJ...896L..18S} speculated that these small flux-ropes can manifest as an outward propagating Alfv¨¦nic fluctuation which might be responsible for the widespread switchbacks observed by Parker Solar Probe (PSP) in solar wind (SW). These observations support that the SIMFRs, at least in part, originate from the solar corona like MCs.


On the other hand, \citet{2000GeoRL..27...57M} hold that in contrast with MCs, SIMFRs originate from interplanetary multiple reconnection process at the heliospheric current sheet (HCS). \citet{2008JGRA..113.9105C} found that the size, proton temperature, and the expansion rate distributions of flux-ropes (including MCs) in the SW appear to be discontinuous and bimodal. These findings suggested different source mechanisms for SIMFR and MCs, but the databases used are somewhat different from \citet{2007JGRA..112.2102F}. Moreover, the duration of SIMFRs shows the power-law distribution \citep[][]{2018ApJS..239...12H,2019ApJ...881...58C}, whereas MCs presents a Gaussian-like distribution \citep{2014SoPh..289.2633J}. Also, the radial scale sizes of many SIMFRs are approximately equal to the estimated HCS thickness. Besides, during the solar minimum period, HCSs are less distorted, simultaneously SIMFRs were observed more frequently in the vicinity of the HCSs \citep{2016JGRA..121.5005Y}. These arguments support that the SIMFRs may be derived from the HCSs.

In addition to the two views above, \citet{2015JGRA..12010175F} suggested that SIMFRs can be divided into two categories in terms of whether they are in the vicinity of the HCSs. In recent years, more and more authors inclined to believe that SIMFRs could be produced both in the SW and on the Sun \citep[e.g.,][]{2010ApJ...720..454T,2011ApJ...734....7R,2014SoPh..289.2633J,2016JGRA..121.5005Y,2018JGRA..123.7167H,2020ApJ...894..120M}.


To support any of these views, a classified comparative study on the characteristic attributes of the two populations of SIMFRs is necessary. A critical problem is how to distinguish the SIMFRs from the Sun (denoted by SIMFR-SUN) and from the SW (denoted by SIMFR-SW). We addressed this issue by setting two criteria. One is to determine if there are counter-streaming suprathermal electrons (CSEs) inside a SIMFR. The CSEs, specifically, strahl and halo electrons, are intense beams of suprathermal electrons aligned to the magnetic field, shedding light on the heliospheric magnetic topology (e.g., \citet{1987JGR....92.8519G}). They are frequently observed along magnetic field lines inside MCs. The existence of CSEs inside MCs usually indicate that the MCs root in the Sun and the field lines still are closed \citep{2000JGR...10527261S}. Similar to MCs, $75\%$ of SIMFRs contain CSEs, which probably implies that such SIMFRs originated from the Sun and keep rooting in the Sun at both ends \citep{2015JGRA..12010175F}. Therefore, it is very likely that picking out the SIMFRs with the CSEs inside enables us to obtain the SIMFR-SUN samples. The other criterion is to determine whether SIMFRs are observed in the vicinity of HCSs \citep{2018JGRA..123.7167H}. The SIMFRs close to HCSs have a high probability of originating from the reconnection at HCSs, and vice-versa. In general, judgement was made based on these two criteria, that is, SIMFR-SUN should posses CSEs and stay away from HCSs, while SIMFR-SW should be observed in the vicinity of the HCSs without containing CSEs.



The paper is organized as follows: Section \ref{sec:data} describes data, events selection, and fitting model, Section \ref{sec:method} describes a approach for deriving the internal parameter distributions, and Section \ref{sec:results} presents the \textcolor{red}{comparative} results, followed by a discussion and conclusion in Section \ref{sec:conclusion}.

\section{ Sample Events selection, data description, and model fitting} \label{sec:data}
Since the solar wind ion composition spectrometer (SWICS) onboard the Advanced Composition Explorer (ACE) spacecraft can provide the complete composition data of the SW, our SIMFR events mainly come from the ACE spacecraft observation events in the SIMFRs database (http://fluxrope.info) which was built with the Grad-Shafranov reconstruction technique \citep{2017JPhCS.900a2024Z,2018ApJS..239...12H}. Moreover, considering the scale of the SIMFRs, there is little difference between WIND and ACE observations in most cases, SIMFRs in \citet{2016JGRA..121.5005Y} observed by WIND spacecraft were therefore employed as supplements. Subsequently, the ACE data were visually checked according to the criterion given by \cite{2007JGRA..112.2102F}.

Alfv¨¦nic fluctuation in SW may show similar magnetic field features with SIMFRs. Thus they could be easily mistakenly considered as SIMFRs \citep{2010ApJ...720..454T,2010AIPC.1216..240M,2010JGRA..115.8102C}. In the database we used, alfv¨¦nic fluctuations have been removed \citep{2016JGRA..121.5005Y,2017JPhCS.900a2024Z}. As the charge-state and elemental-abundance data of ACE/SWICS have the cadence of 1 hr and 2 hrs respectively, it is more reasonable to choose the SIMFRs with long duration for analysis and we picked up the SIMFRs of duration between 3 - 12 hrs from the database. Subsequently, the candidate SIMFRs were cross-checked by fitting with the constant-$\alpha$, force-free, cylindrically symmetric flux-ropes model \citep{1950Ark.35...361, 1988JGR....93.7217B,1990JGR....9511957L}, which is simple but still widely used to date. A candidate was selected when its boundaries were close to the best-fit boundaries and the fitting results were acceptable, e.g., normalized root-mean-square $\chi_{n}<0.6$ \citep{2018JGRA..123.3238W}. In addition, because the edge measurement to an flux-rope brings a large error to the fitting results and may misidentify an MC as an SIMFR, the impact parameter ($d$), i.e., the closest distance of the spacecraft to the rope axis, should satisfy $d\leq0.7$, in units of flux-rope radius ($R_{c}$).

For the first SIMFRs' classification criterion, we check the CSEs status inside the qualified SIMFRs, using the same criterion as \citet{2018JGRA..123.7167H}, i.e., the percentages of CSEs intervals covering more (less) than $10\%$ of the SIMFRs, is a (no) signature of CSEs in the SIMFRs. For the second criterion, the vicinity of the HCSs is defined as $\pm 3$ days around the HCSs \citep[e.g.][]{2015JGRA..12010175F} from the HCS lists published by Leif Svalgaard (http://www.leif.org/research/sblist.txt).

We identified and modelled SIMFRs observed by the ACE spacecraft from February 1998 to August 2011, before the ACE/SWICS have been recalibrated due to hardware anomaly on August 23 2011. The magnetic field strength data were provided by ACE/MAG every 4-mins. SW bulk speed, proton temperature, and Helium-to-proton density ratio ($\mathrm{He/P}$) data were provided with a cadence of 1-hr by ACE/SWEPAM. The rest of the data came from ACE/SWICS, in which the charge-state of C, O, Mg, Si, Fe, $\mathrm{O^{7+}/O^{6+}}$, $\mathrm{C^{6+}/C^{4+}}$, $\mathrm{C^{6+}/C^{5+}}$, $\mathrm{Fe/O}$ used 1-hr cadence data, proton number density used 12-mins cadence data, and the Ne/O, Mg/O, Si/O, C/O, He/O used 2-hrs cadence data. The identification of CSEs in SIMFRs used the plasma ions and electrons data measured by 64 s ACE/SWEPAM measurements, as well as 24 s WIND/3DP and 12 s WIND/SWE measurements. The electrons energy analysis in this study range approximately from 80 eV to 300 eV, the same as \citet{2015JGRA..12010175F}.

\begin{figure}[t]
\centering
  \begin{tabular}{@{}cccc@{}}

    \includegraphics[width=.45\textwidth]{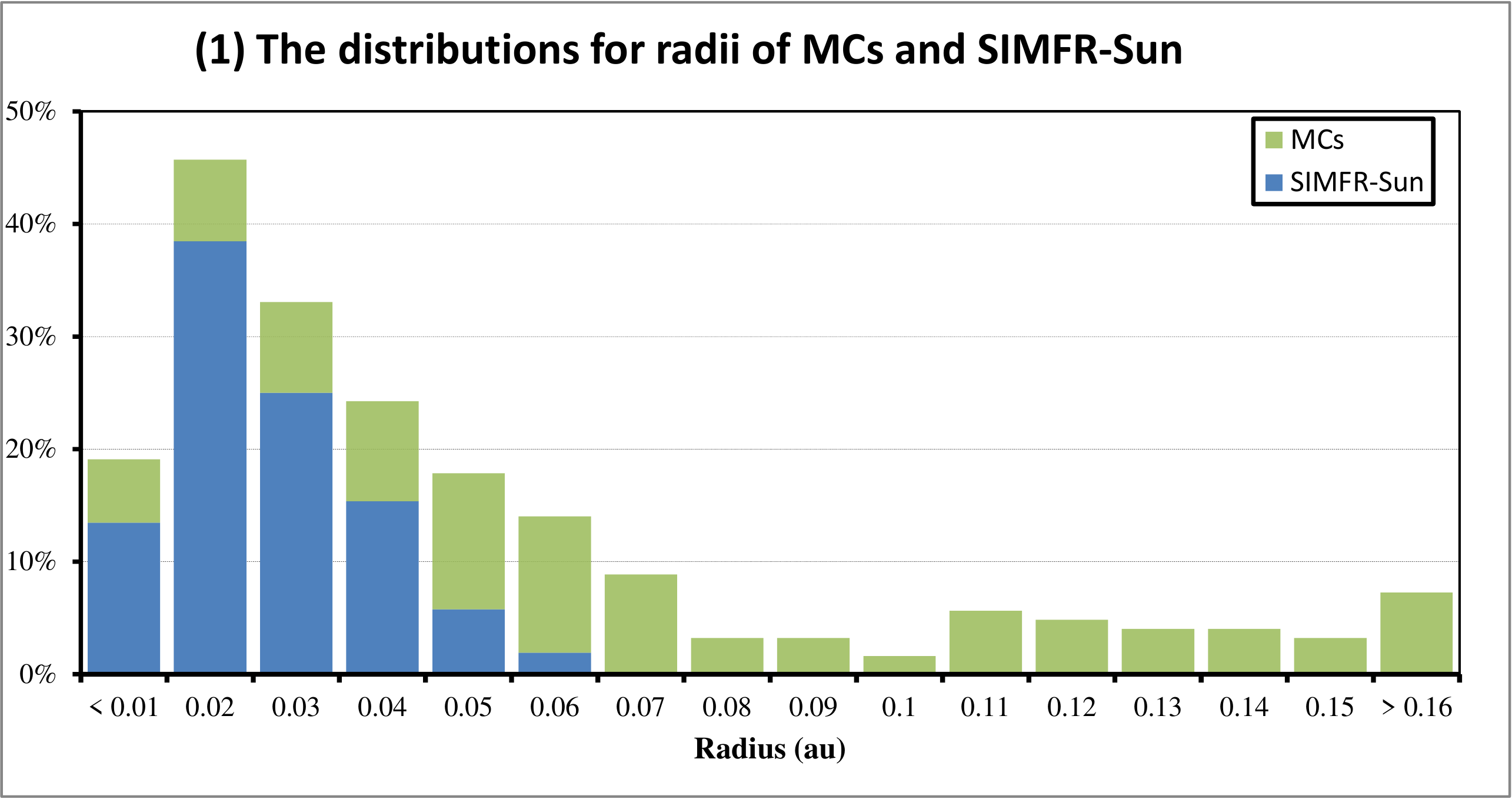} &
    \includegraphics[width=.45\textwidth]{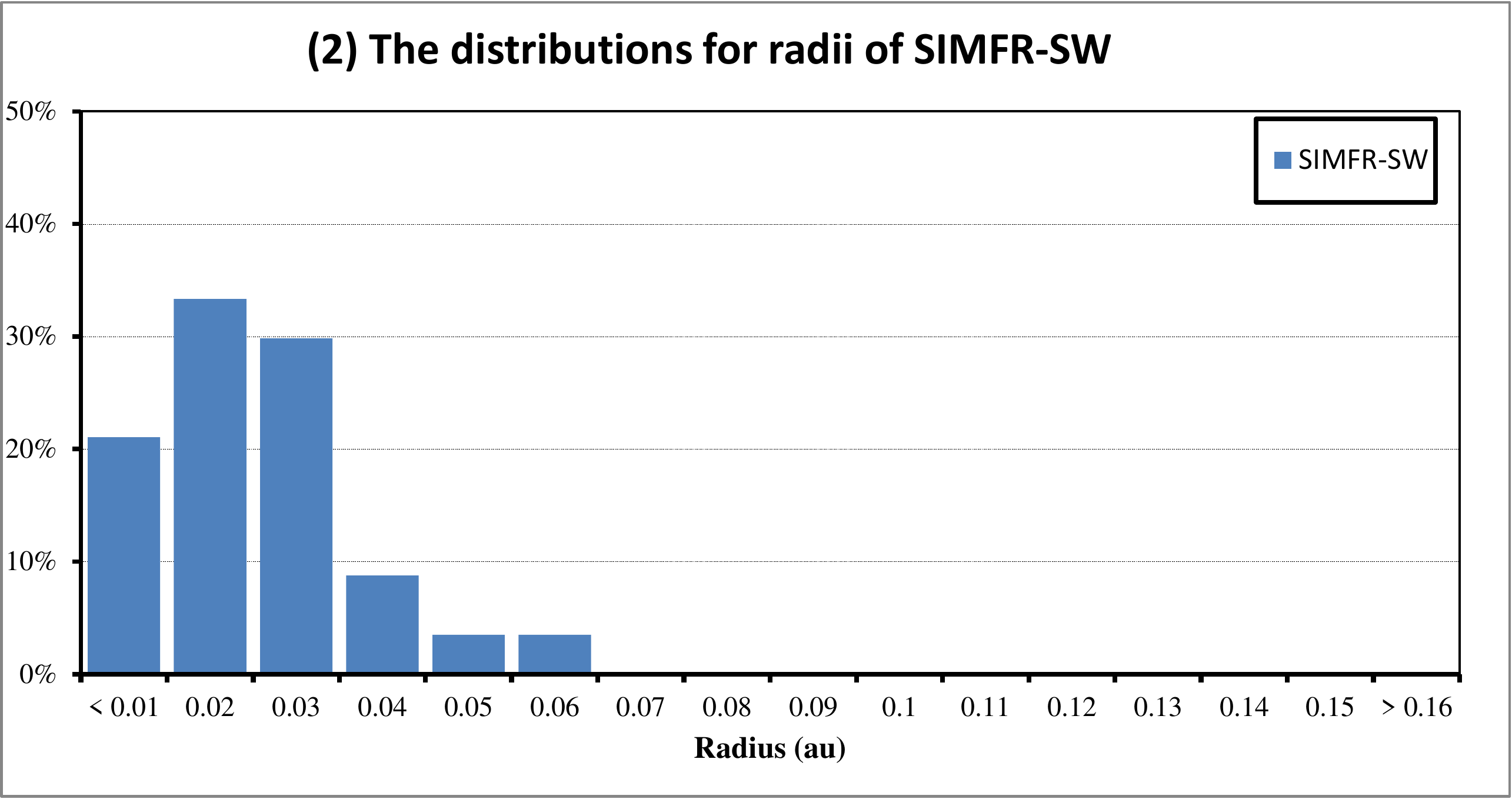} \\

  \end{tabular}
  \caption{The scale distributions of SIMFRs and MCs by proportions.
}
  \label{fig:radii}
\end{figure}

Finally, 52 SIMFR-SUN events and 57 SIMFR-SW events were selected. The events and fitting results are posted at http://sss.ynao.ac.cn/Public/upload/file/simfr.pdf. For MC samples, we used a total of 124 events and the fitting result of \citet{2020ApJ...893..136H}, whose fitting model and procedure are the same as this study. The scale distributions of all the flux-ropes by proportions are shown in Figure \ref{fig:radii}. As can be seen in Figure \ref{fig:radii} (1), SIMFR-SUN and MCs are roughly continuous in scale distribution. Note that if we take the shorter interval SIMFRs (less than 3 hr) into account, the proportion of $< 0.1$ au will become much higher. The comparison between Figure \ref{fig:radii} (1) and (2) show that SIMFR-SW tend to be smaller in scale than SIMFR-SUN. It is consistent with the finding in \citet{2019AGUFMSH43D3372X} that SIMFRs in the surrounding SW have a shorter duration than SIMFRs in ICME bodies.

\section{Approach of inferring spatial position} \label{sec:method}


We used \citet{2003JGRA..108.1239L} and \citet{2020ApJ...893..136H}' s approach to extract a normalized position (denoted as $x$) for every measured quantity.  The measured quantity has been coupled with the radial distance inside the cylinder using the model geometry.
\begin{equation}\label{eq:normalized}
  x = \frac{\mathrm{|OA|}}{R_{c}}
\end{equation}

In the cross-section of the cylindric flux-rope, $x$ can be derived from Eq.(1). In this equation, $\mathrm{|OA|}$ is the projection-distance from the axes of the flux-rope to the measurement position. It can be calculated by $R_{c}$, $d$, and the spacecraft travelled projection-distance inside the flux-rope. By applying this approach, we can obtain the normalized structure among different sizes of flux-ropes and construct the statistical average of any measured quantity.

\section{Results} \label{sec:results}

\begin{figure}[b]
\centering
  \begin{tabular}{@{}cccc@{}}
    \includegraphics[width=.45\textwidth]{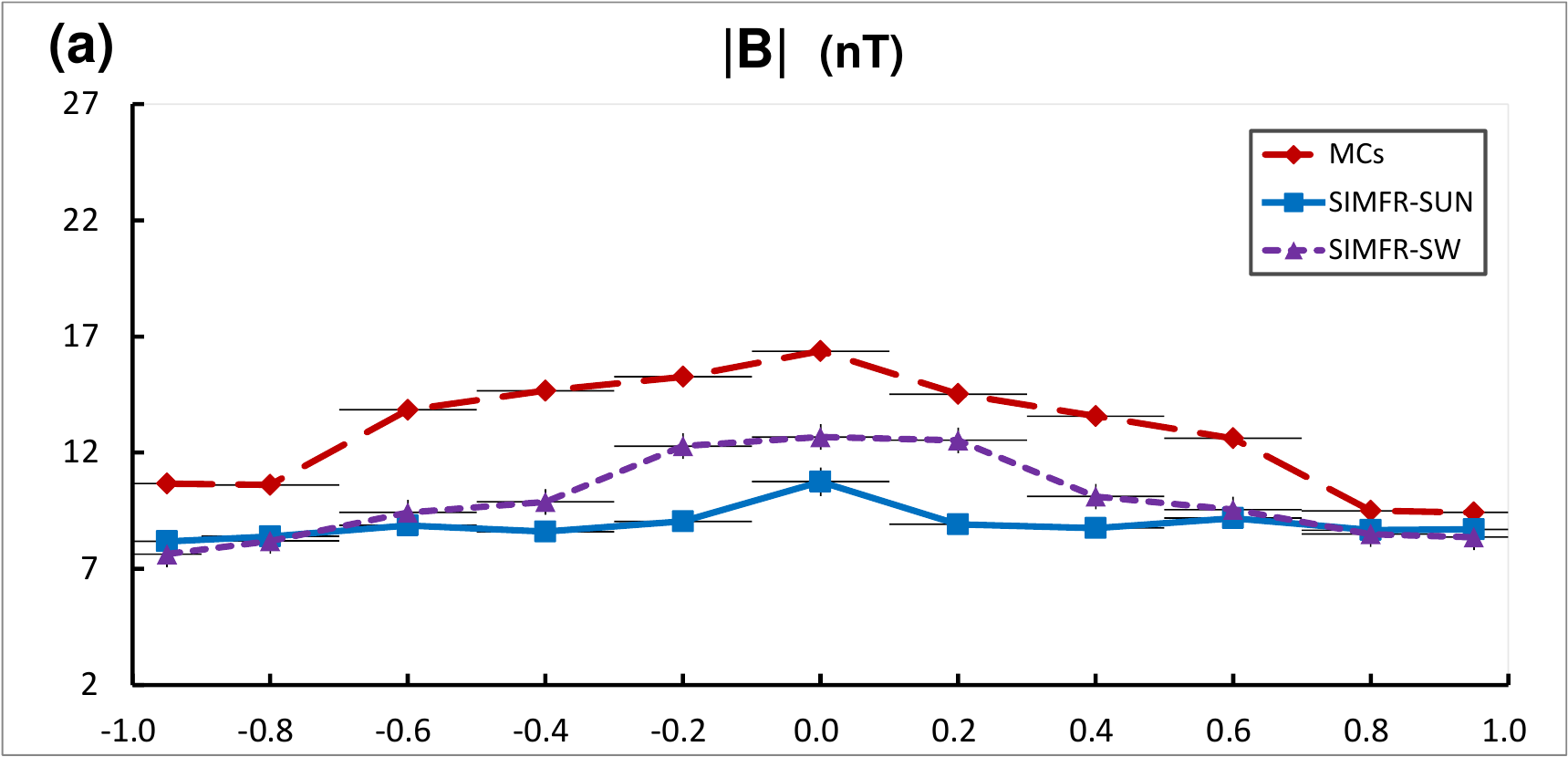} &
    \includegraphics[width=.45\textwidth]{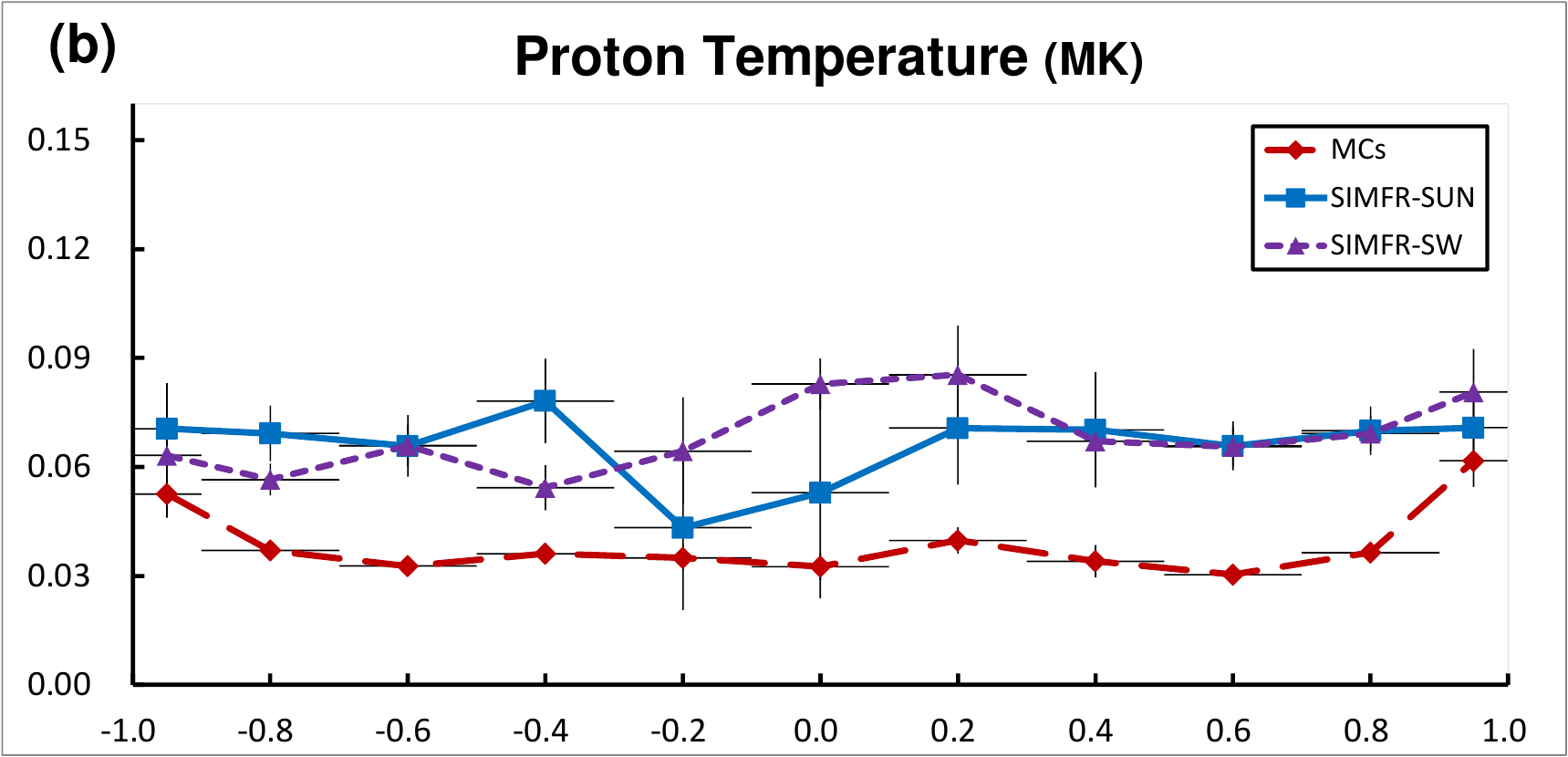}  \\
    \includegraphics[width=.45\textwidth]{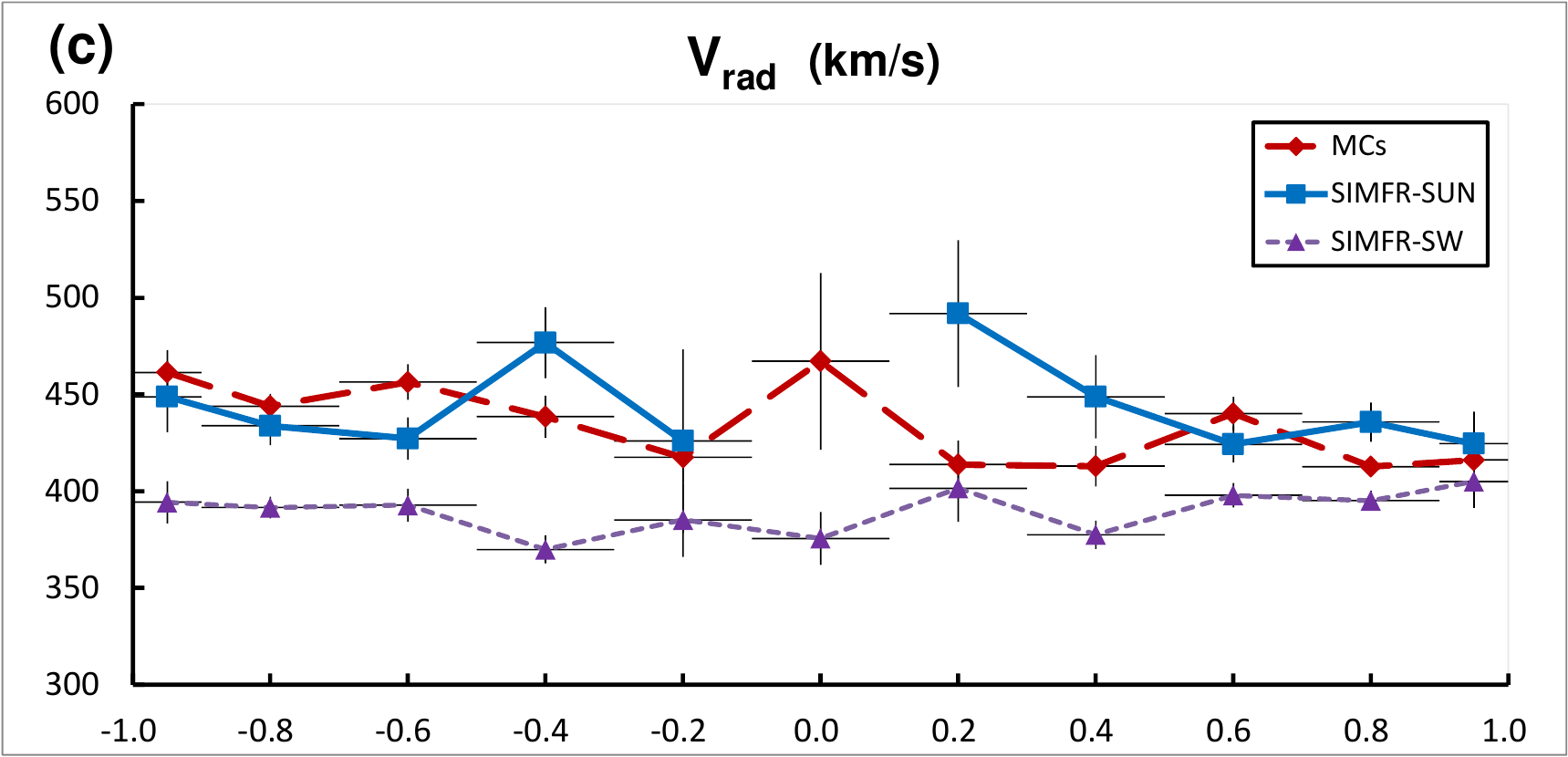} &
    \includegraphics[width=.45\textwidth]{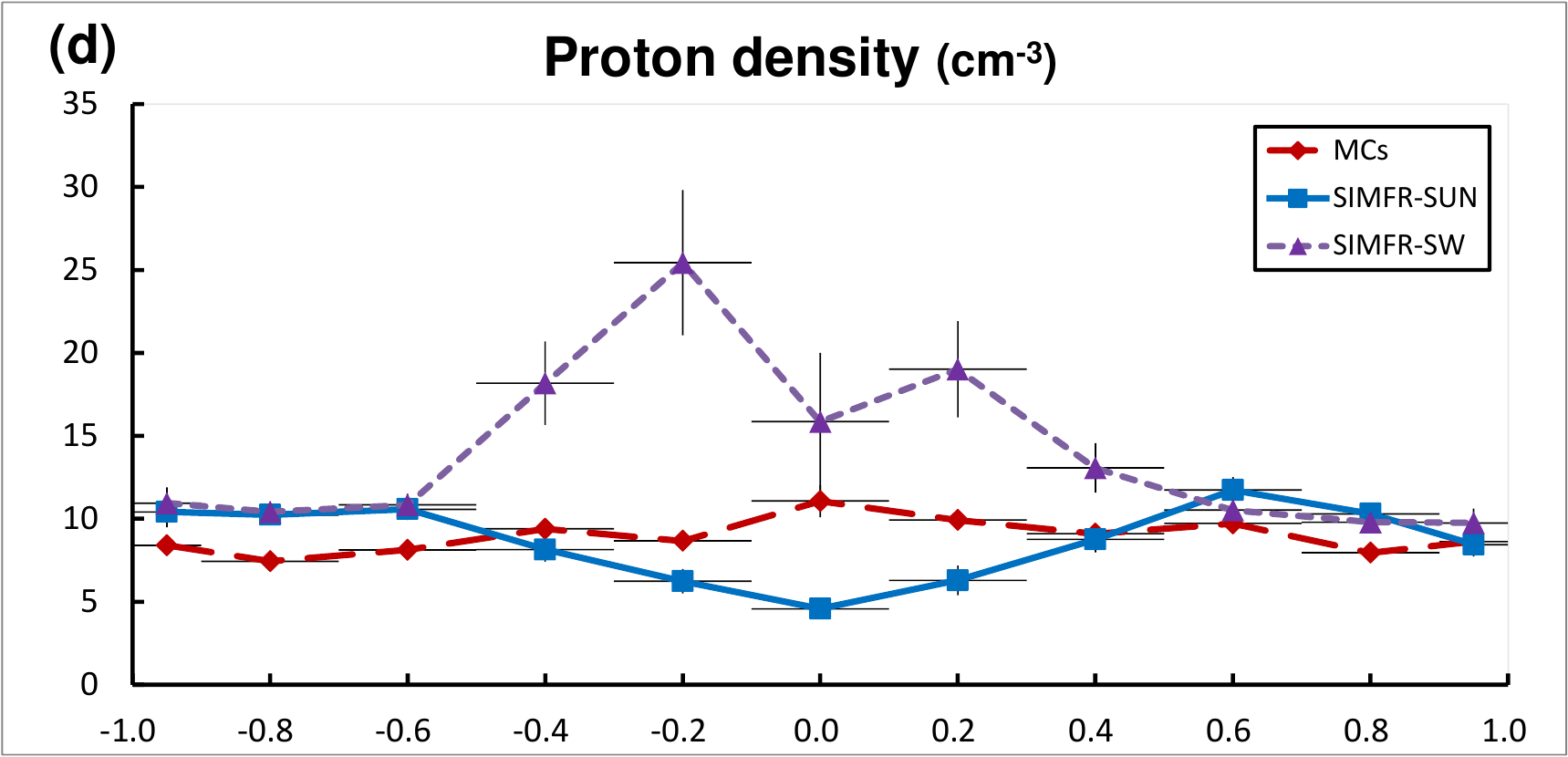}  \\
    \includegraphics[width=.45\textwidth]{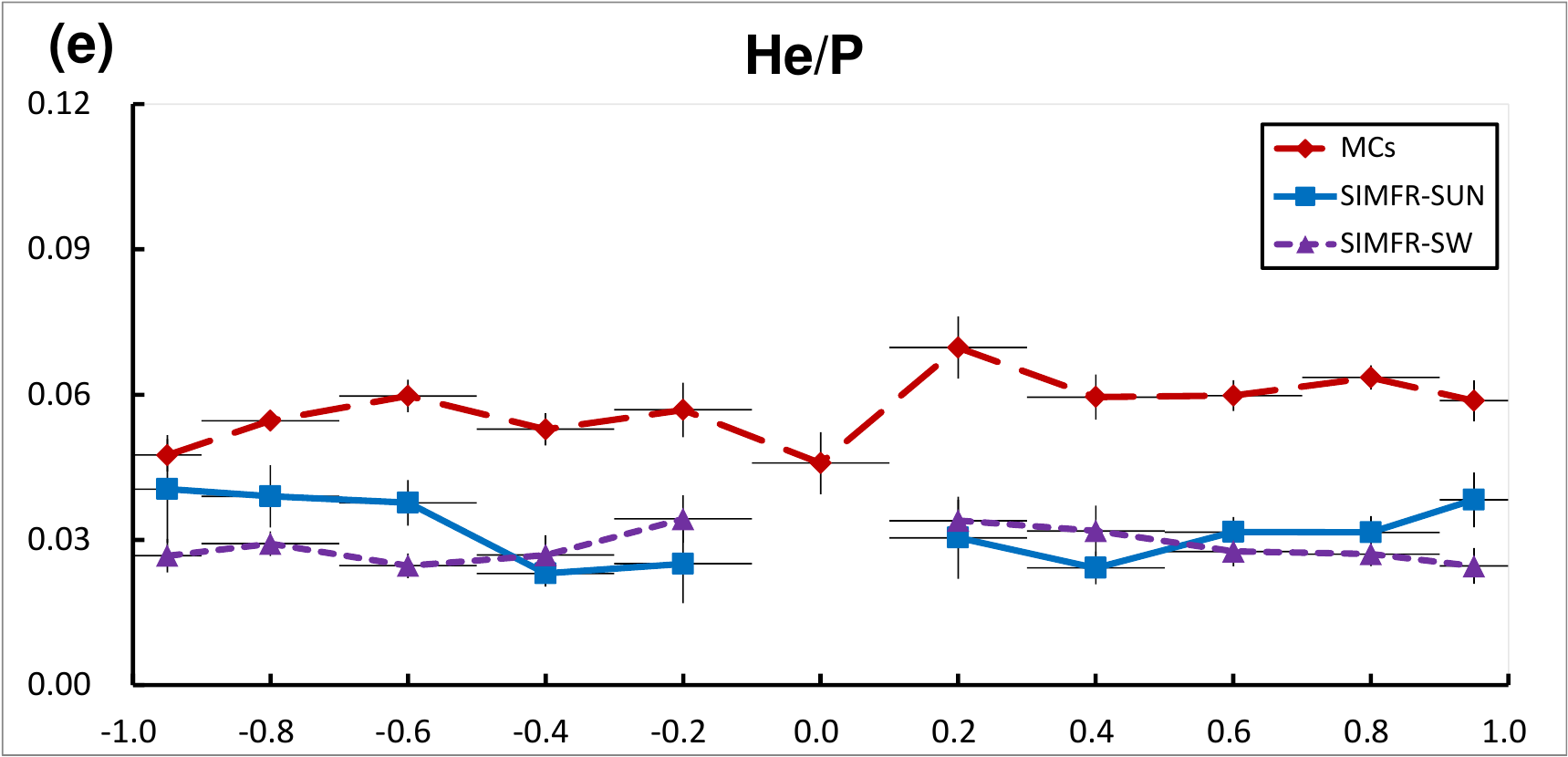} &
    \includegraphics[width=.45\textwidth]{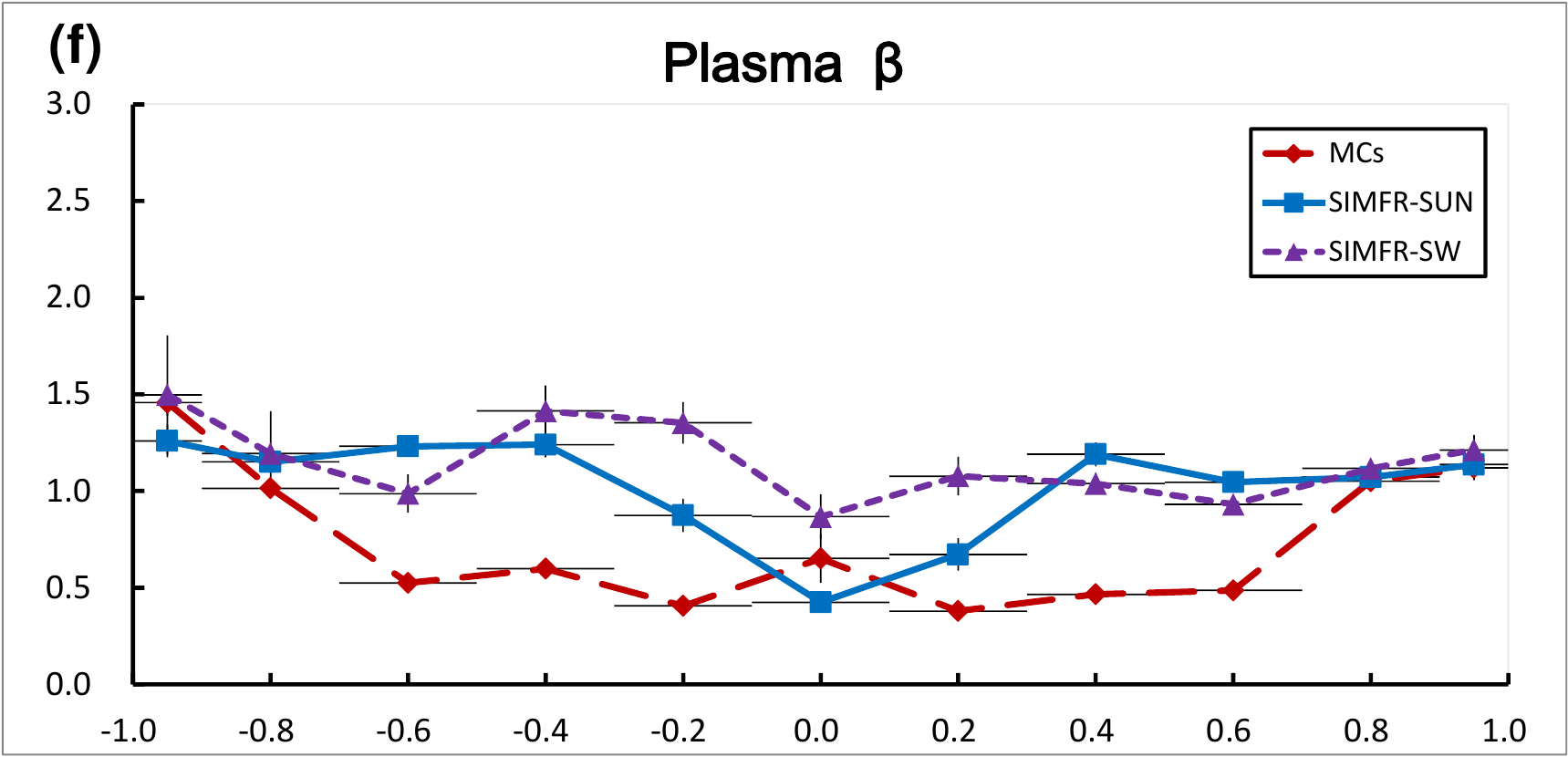} \\
 \end{tabular}
  \caption{Statistical distributions of magnetic field magnitude and plasma parameters inside flux-rope by ACE during 1998 - 2011. The blue, purple, and red lines denote SIMFR-SUN, SIMFR-SW and MCs respectively. The vertical error bars represent the standard error of the average in each bin. Some values in the center were excluded due to small sample size (less than 5).
}
  \label{fig:plasma}
\end{figure}

    \begin{figure}[b]
\centering
  \begin{tabular}{@{}cccc@{}}
    \includegraphics[width=.45\textwidth]{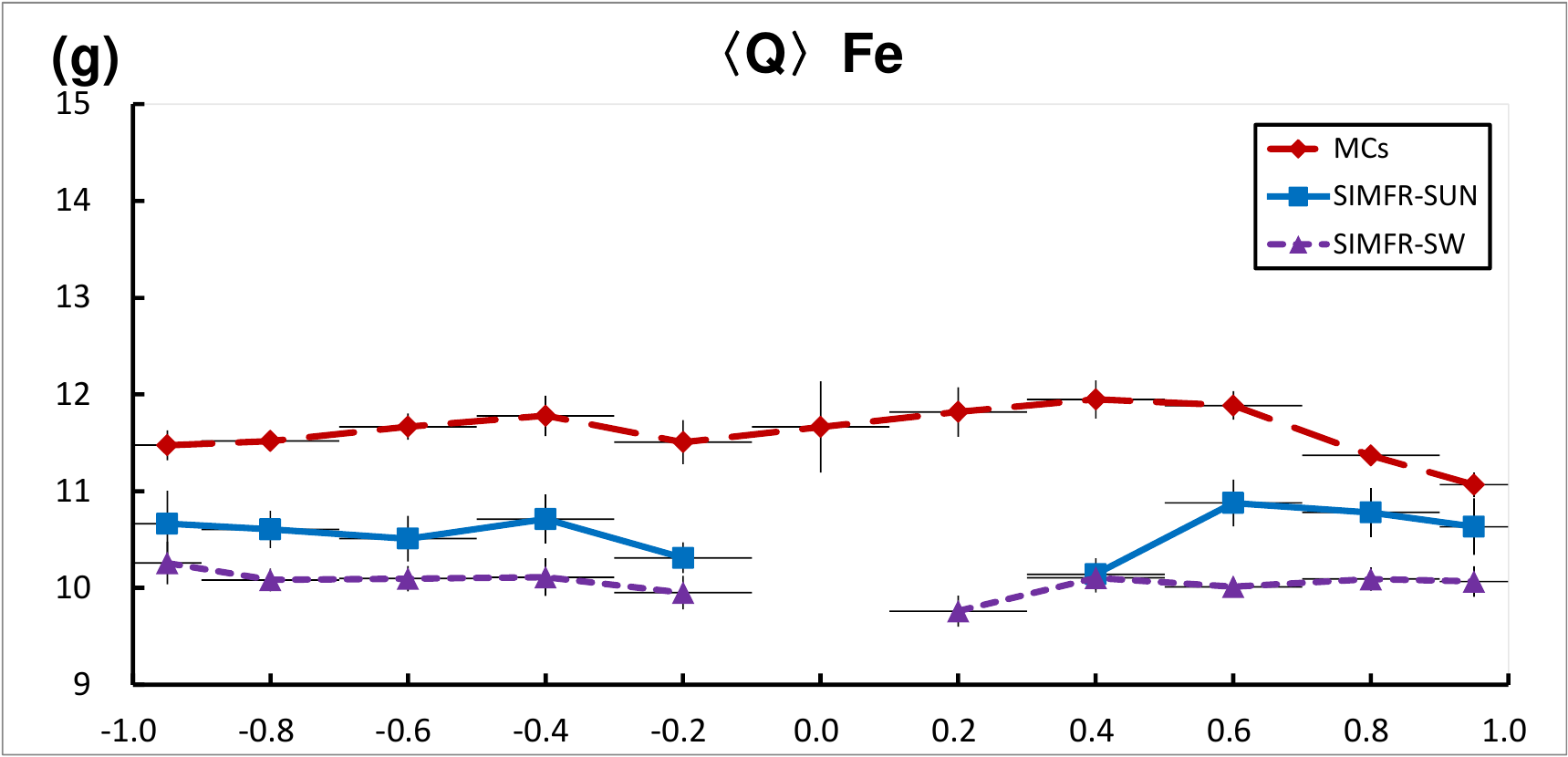} &
    \includegraphics[width=.45\textwidth]{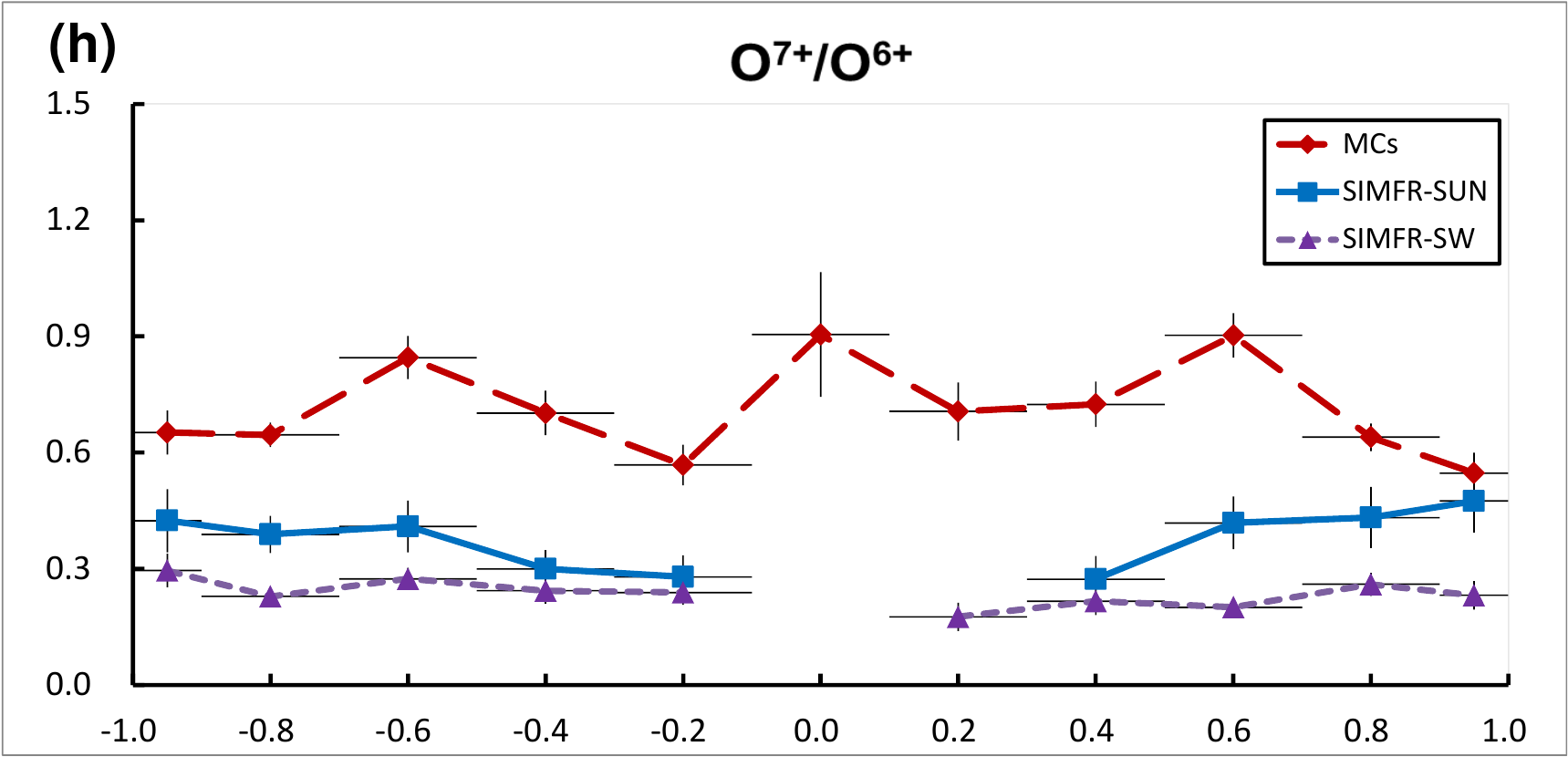} \\
    \includegraphics[width=.45\textwidth]{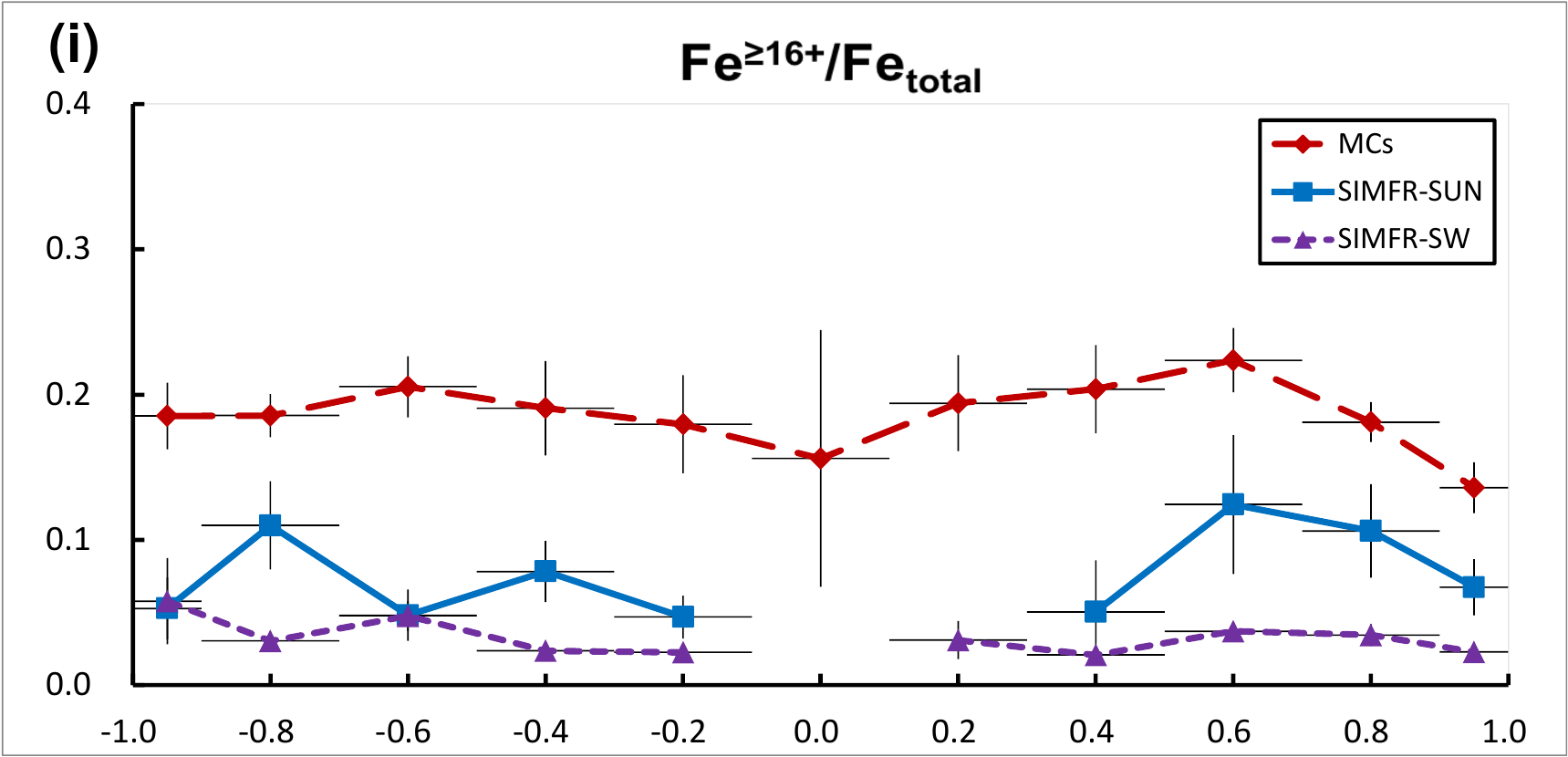} &
    \includegraphics[width=.45\textwidth]{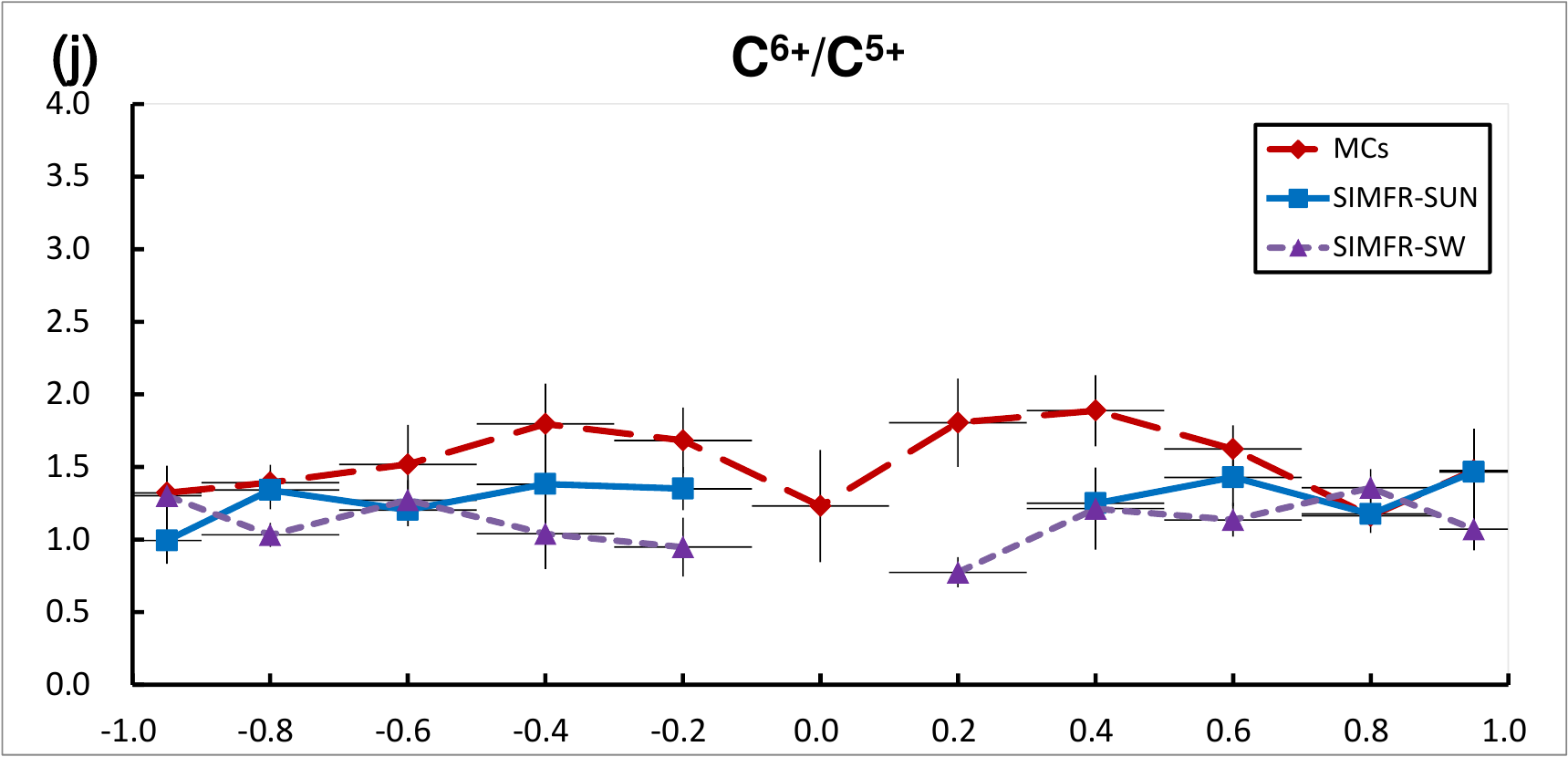}

       \end{tabular}
   \end{figure}

   \begin{figure}[t]
\centering
  \begin{tabular}{@{}cccc@{}}
    \includegraphics[width=.45\textwidth]{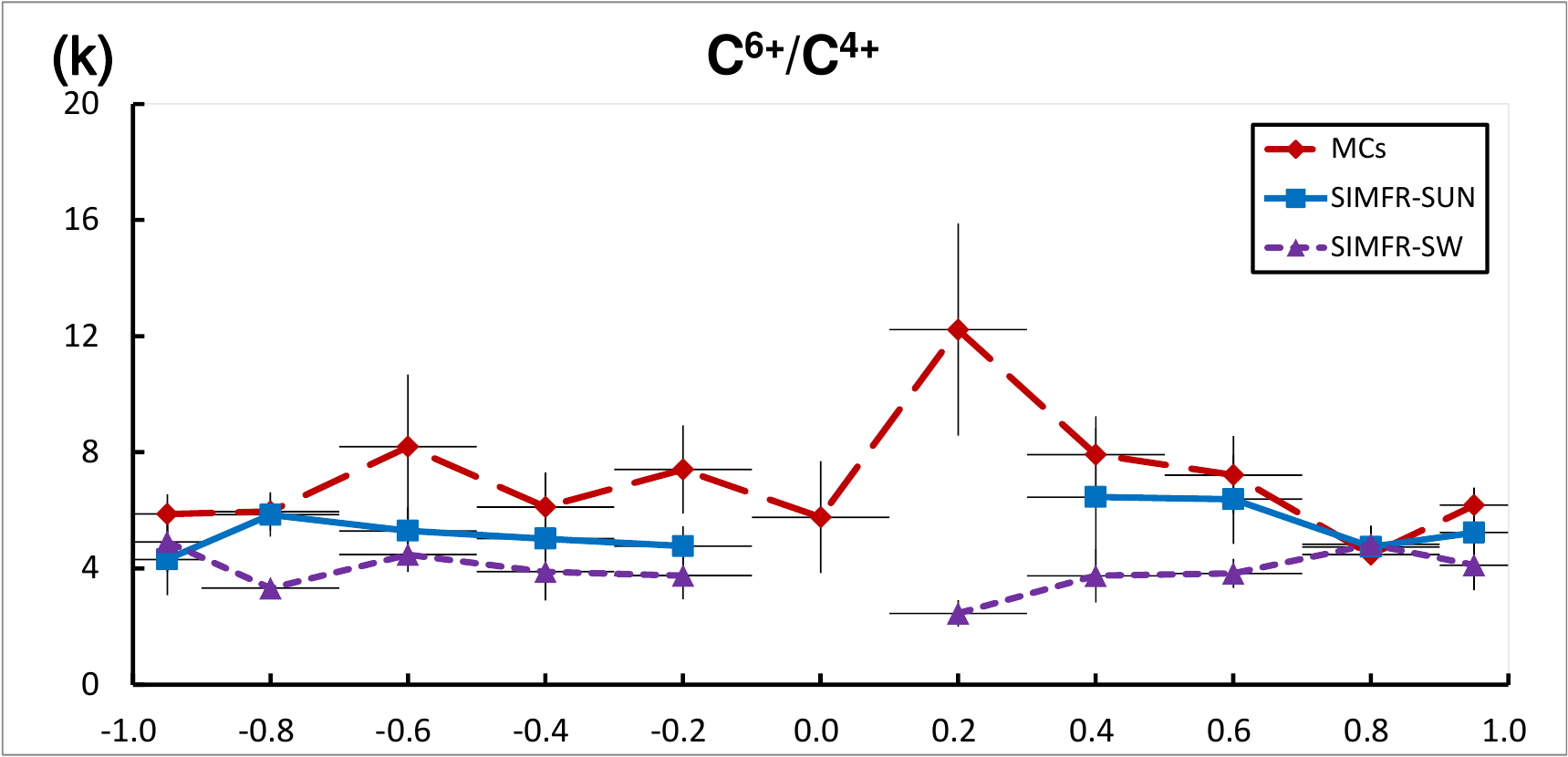} &
    \includegraphics[width=.45\textwidth]{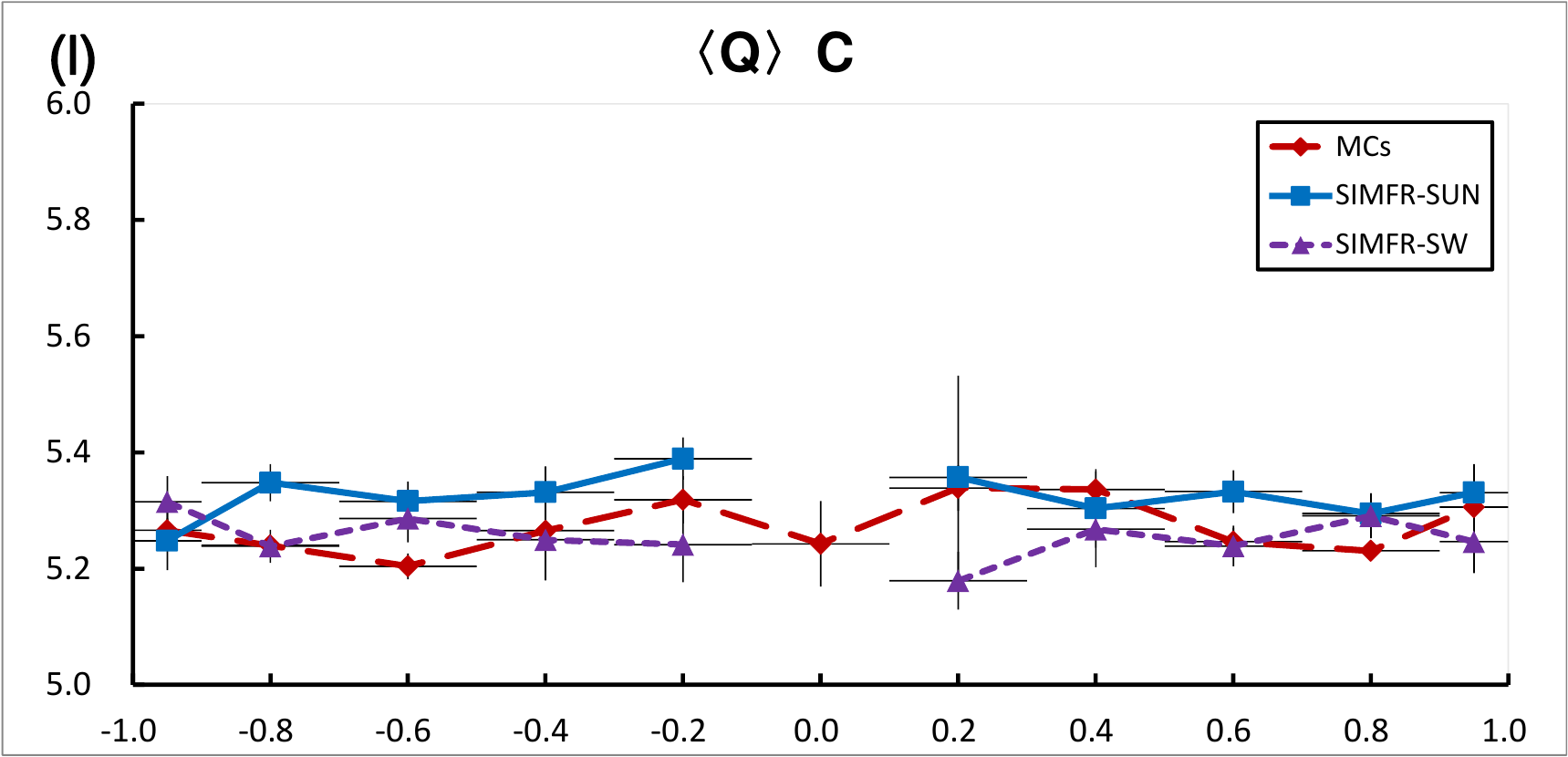} \\
    \includegraphics[width=.45\textwidth]{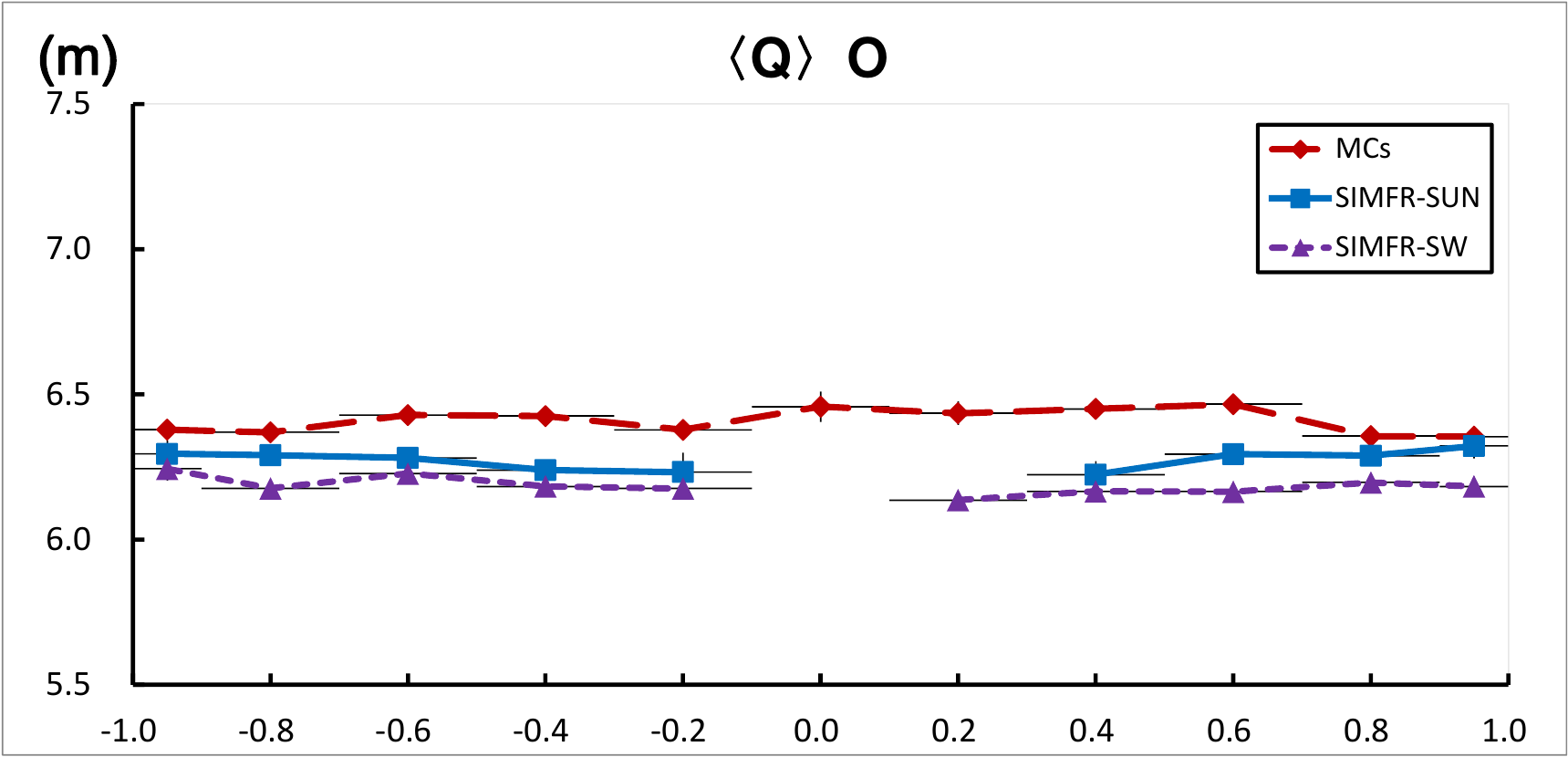} &
    \includegraphics[width=.45\textwidth]{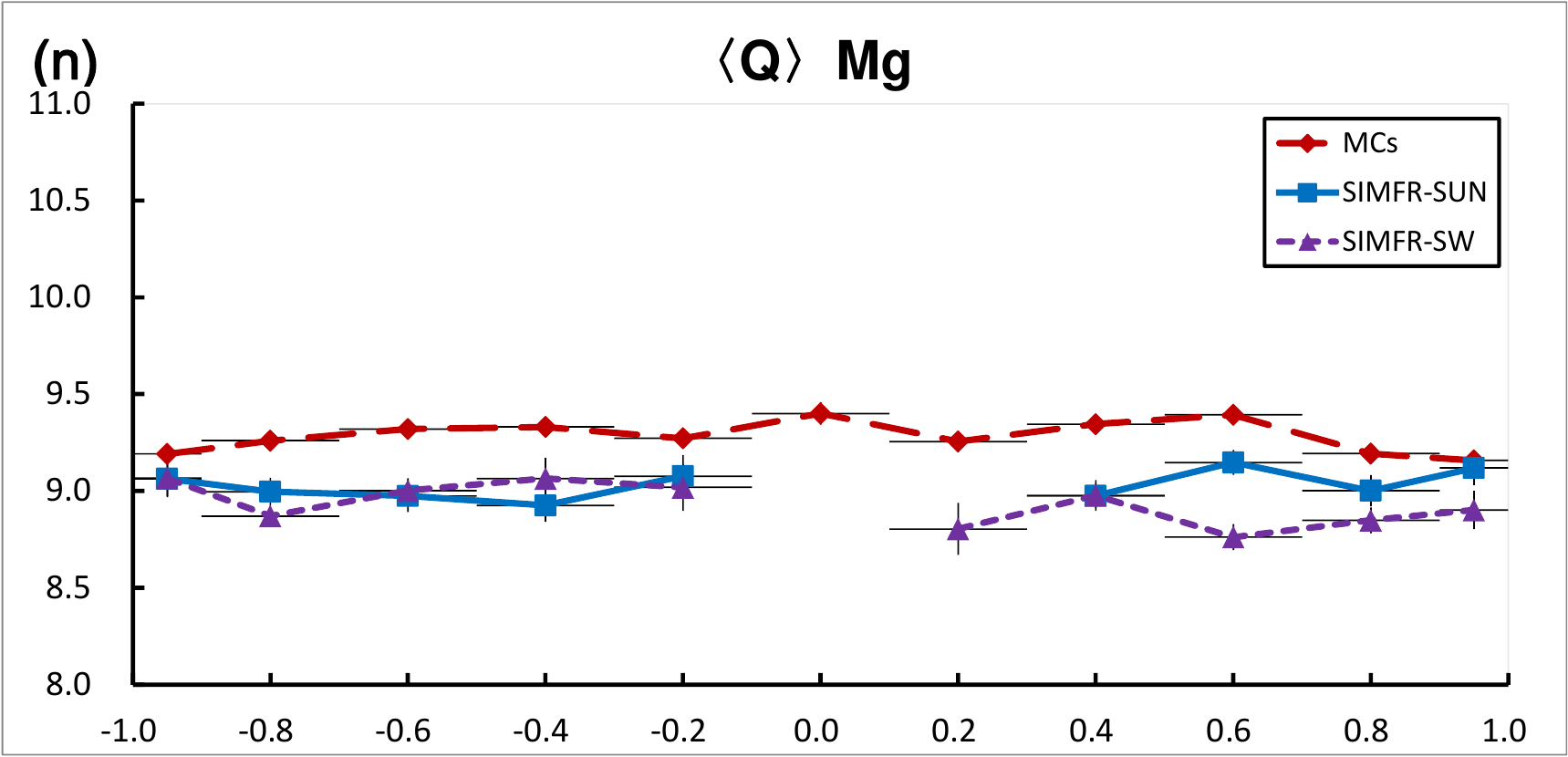}  \\
    \includegraphics[width=.45\textwidth]{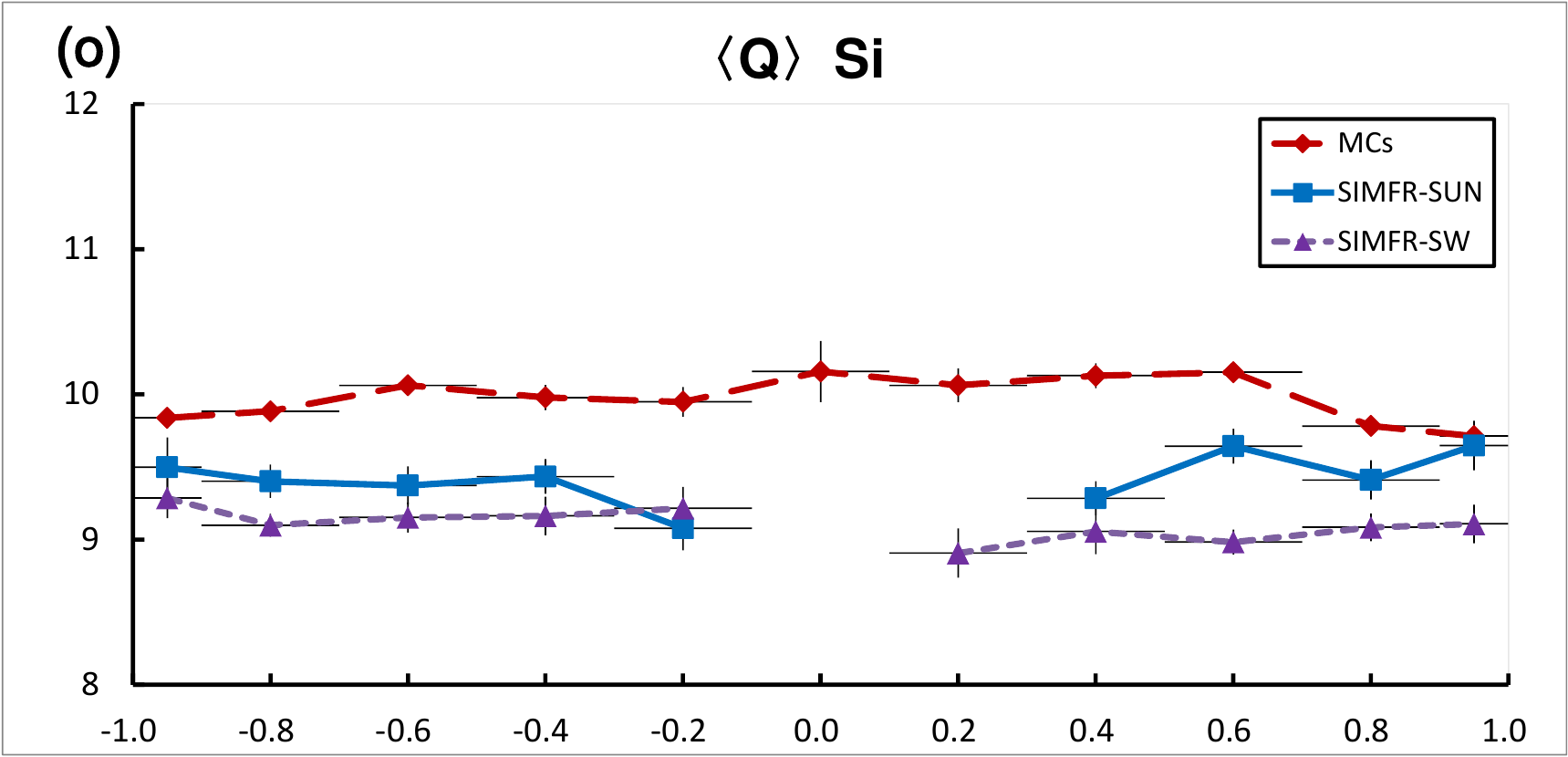} &
  \end{tabular}
  \caption{The same as Figure \ref{fig:plasma}, but for the statistical distributions of ion charge-states inside flux-rope.
}
  \label{fig:chargestates}
\end{figure}

\begin{figure}[t]
\centering
  \begin{tabular}{@{}cccc@{}}

    \includegraphics[width=.45\textwidth]{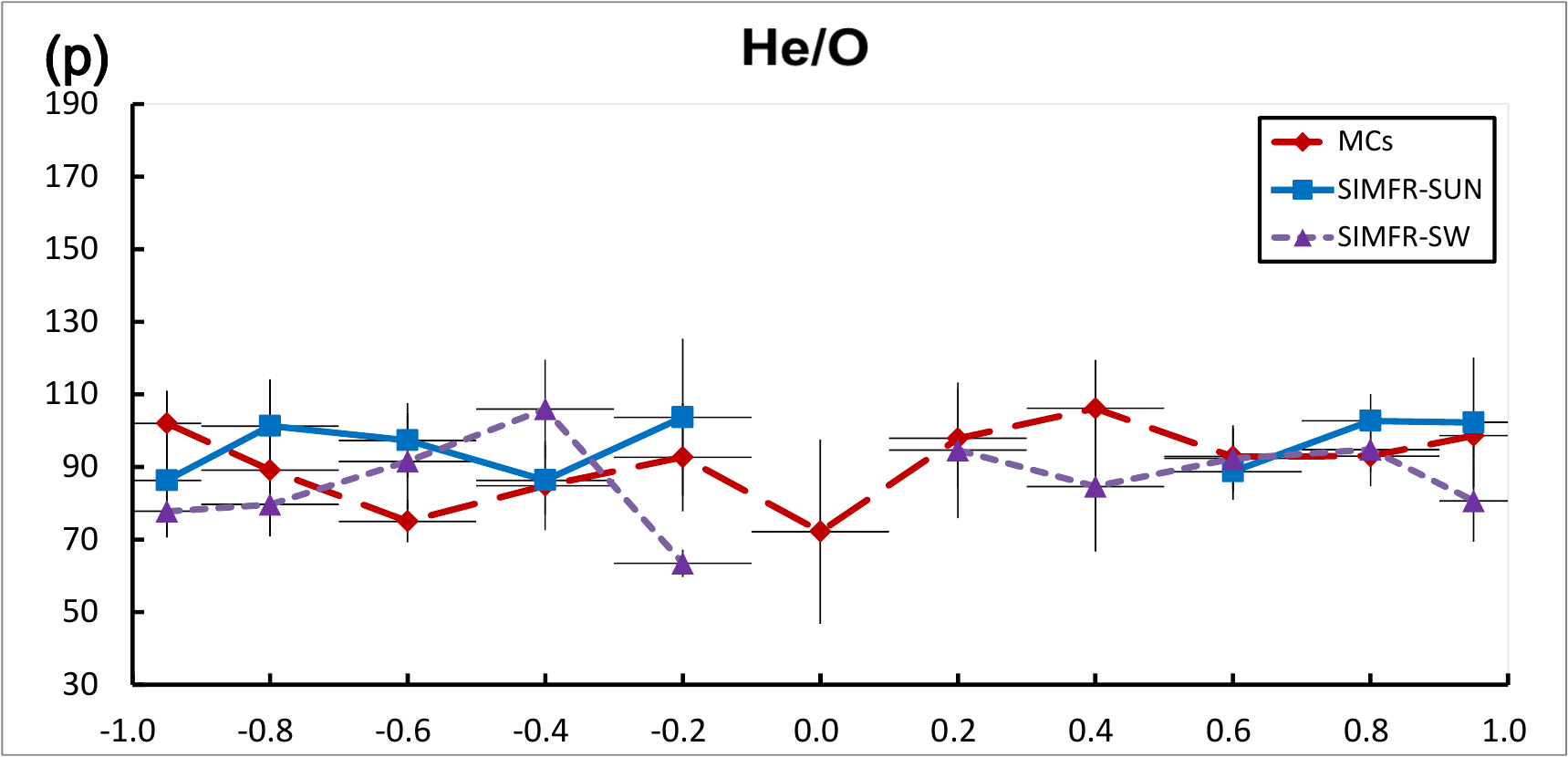} &
    \includegraphics[width=.45\textwidth]{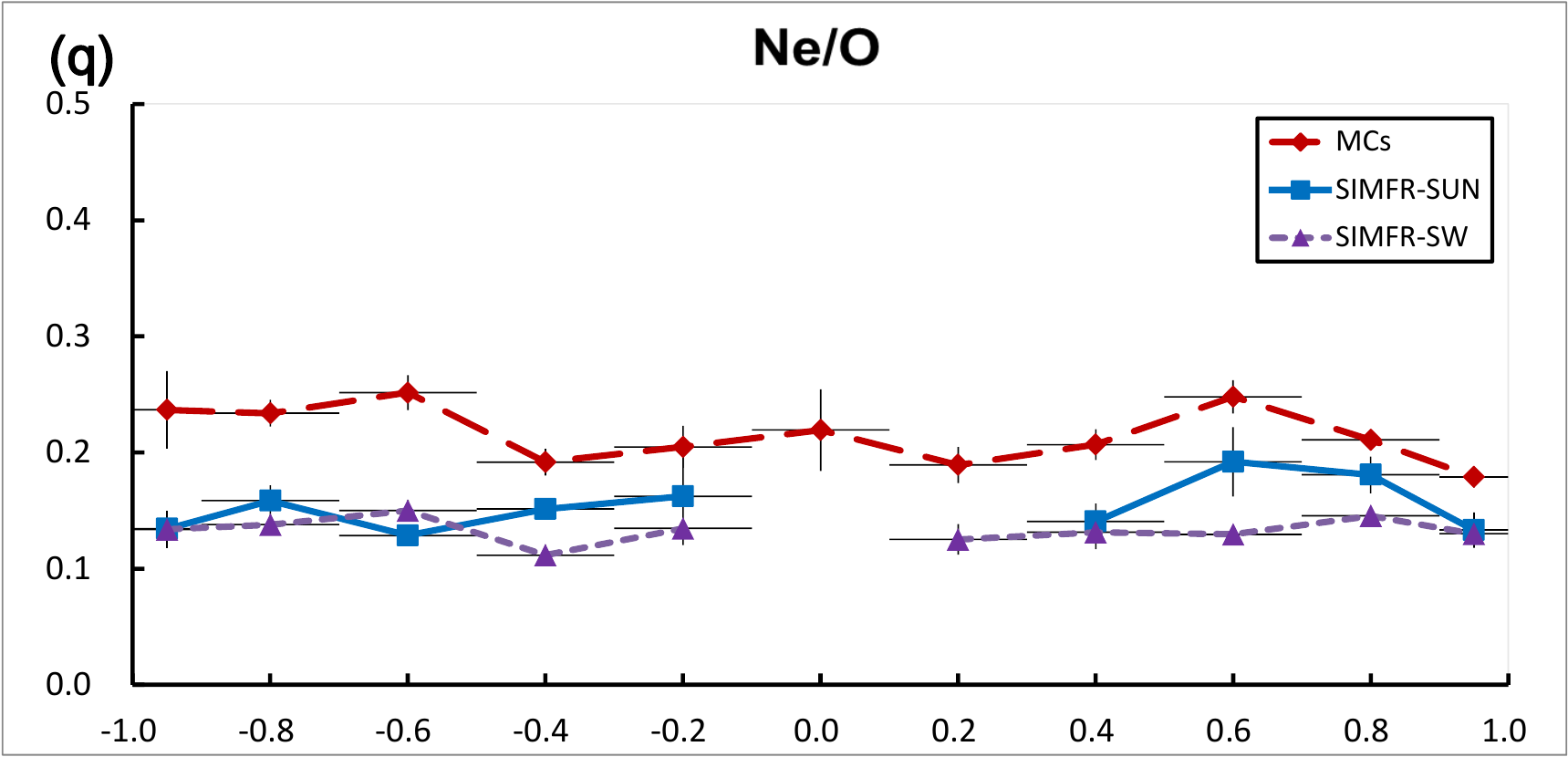} \\
    \includegraphics[width=.45\textwidth]{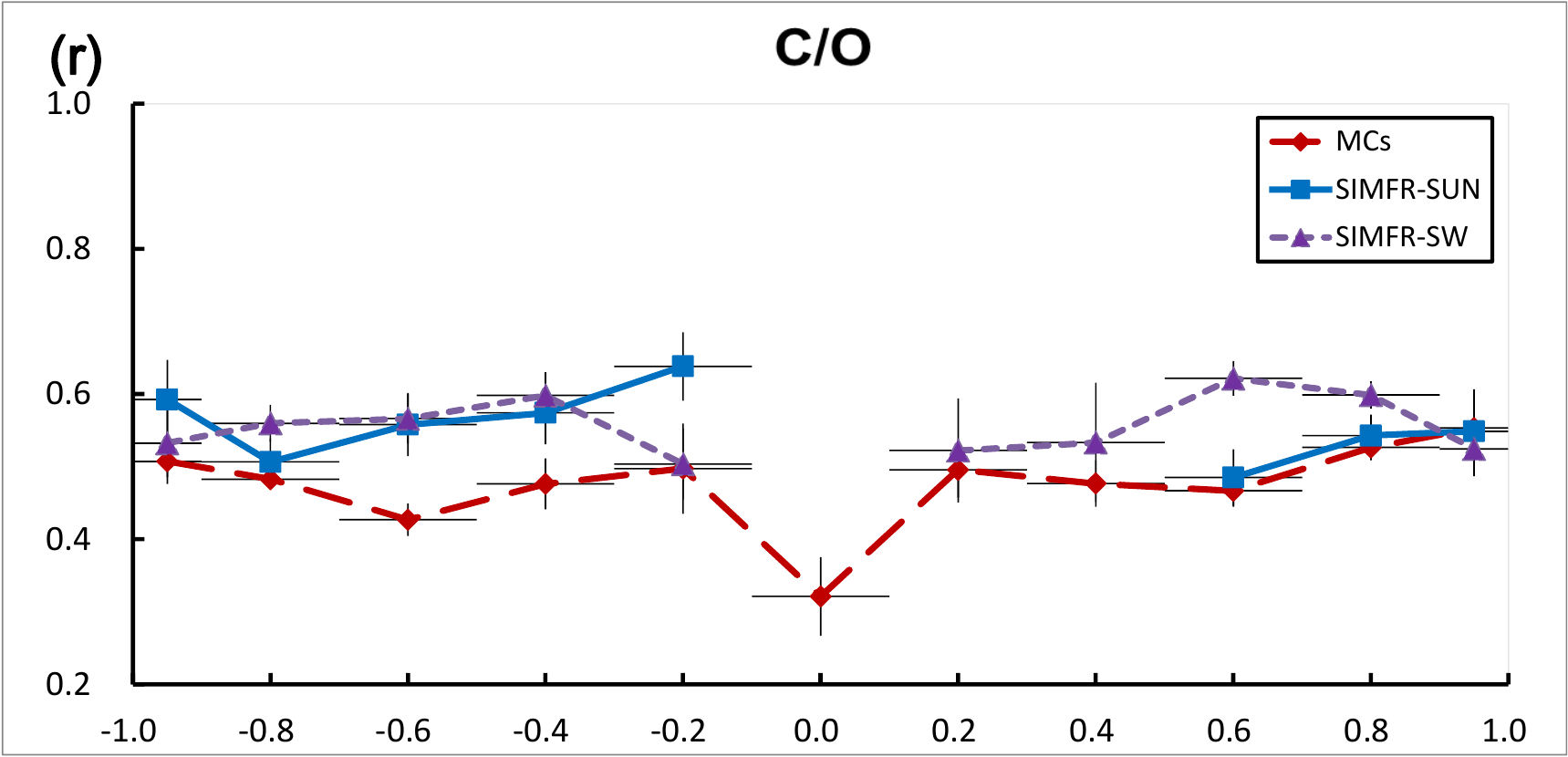} &
    \includegraphics[width=.45\textwidth]{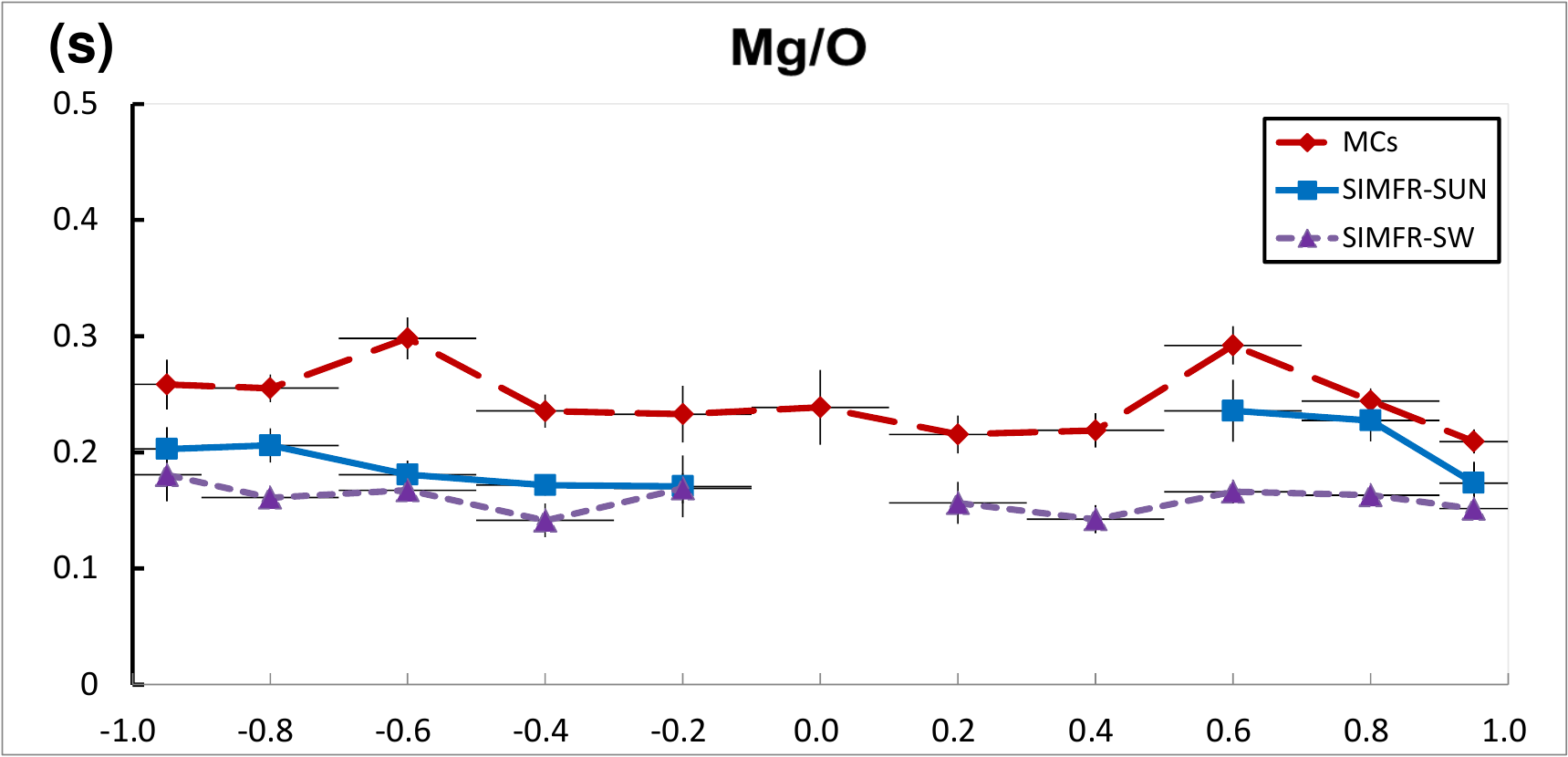}

  \end{tabular}
  \caption{The same as Figure \ref{fig:plasma}, but for the statistical distributions of elemental abundances inside flux-rope.
}
  \label{fig:abundance}
\end{figure}

In this study, we distinguished two populations of SIMFR, i.e. SIMFR-SUN (52 events) and SIMFR-SW (56 events), according to whether it contains CSEs and stays far away from HCSs. Combining MCs (124 events) and the abovementioned two populations produces three populations of flux-rope. Subsequently, the plasma and the composition parameter distributions inside the three popolations were made. Results are shown in Figure \ref{fig:plasma} to \ref{fig:abundance} , which has linked all the measured quantities to the corresponding normalized position. The normalized space was divided into 11 bins, the average value in each bin was calculated, and the error bars denote the standard errors. When the ACE spacecraft passes through flux-ropes, the parameters distributions inside flux-ropes (Figure \ref{fig:plasma} to \ref{fig:abundance}) will appear from left to right. In other words, the negative $x$-axis is the earthward side, whereas the positive $x$-axis is the sunward side. To ensure the credibility, we removed the values of bins with less than 5 samples. The related explanations of every distribution are as follows:

Figure \ref{fig:plasma} is about the magnetic field and basic plasma parameter distributions. Each panel is described as follows:

(a) Magnetic field magnitude ($\left|B\right|$). All of the 3 populations of flux-rope show a domed-like profile. SIMFRs have lower $\left|B\right|$ than MCs, but more symmetry. SIMFR-SW have a higher central value than SIMFR-SUN in the center.

(b) Proton temperature ($T_{p}$). SIMFRs show higher $T_{p}$ than MCs. SIMFR-SUN and SIMFR-SW are comparable. Besides, MCs show hotter $T_{p}$ on edges which may be caused by the errors in MC boundary selections.

(c) Radial velocity of the SW ($V_{rad}$). There is no obvious $V_{rad}$ decrease throughout SIMFRs, implying little expansion for them. The average $V_{rad}$ of SIMFR-SUN is similar to that of MC, obviously higher and more fluctuant than SIMFR-SW.

(d) Proton density ($N_{p}$). SIMFR-SUN and MCs are comparable in $N_{p}$, but SIMFR-SUN tends to be depleted in the center. It is worth noting that the $N_{p}$ of SIMFR-SW are much pronounced than SIMFR-SUN and MCs in the center. In Table 1 of \citet{2020ScChE..63..183F}, SIMFRs were characterized by indistinctively lower density, which should be attributed to the remarkably high density inside a subset of SIMFRs (i.e. SIMFR-SW), according to our results.

(e) He/P ratio (also known as helium abundance). SIMFR-SUN and SIMFR-SW are on the same level, which are much lower than MCs and close to that in the SW in \citet{2020ApJ...893..136H}.

(f) Plasma $\beta$. It was calculated by the ratio of thermal to magnetic pressure. $ \beta = NkT/(B^{2}/8\pi)$, where $N$, $T$, and $B$ are the density ($cm^{-3}$), temperature ($K$) of plasma (protons), and magnetic field strength (nT) respectively. For SIMFRs, SIMFR-SUN tends to be depleted in the center, while SIMFR-SW tends to be flat. The plasma $\beta$ of both SIMFR-SUN and SIMFR-SW roughly equals to 1, which is higher than MCs. In addition, for SIMFR-SUN and MCs, there are anti-correlation between $\beta$ and He/P trend, but this characteristic is not obvious in SIMFR-SW. \citet{2018arXiv180703579Y} also found that such a relation exists in the center of MCs, and is absent in non-MC interplanetary CMEs (Ejecta), which was interpreted as the presence of a helium electric current in MCs. However, the upper limit of the linear size of the helium current was estimated to be only $10\%$ of MC size in their study.

Figure \ref{fig:chargestates} is about ion charge-state ratios and average value distributions:

Panel (g) - (k) are $\mathrm{\langle Q\rangle Fe}$, $\mathrm{O^{7+}/O^{6+}}$, $\mathrm{Fe^{\geq16+}/Fe_{total}}$, $\mathrm{C^{6+}/C^{5+}}$, and $\mathrm{C^{6+}/C^{4+}}$ ratio sequentially. SIMFRs show lower value (ratio) than MCs. For the two populations of SIMFRs, SIMFR-SUN are higher than SIMFR-SW obviously. In addition, MCs display bimodal distributions, and the rear peaks are higher than the front ones, which can be explained by the flare heating and the high-energy electron collisions from the reconnection region \citep{2020ApJ...893..136H}. The higher value (ratio) can also be seen in the rear part of SIMFR-SUN's internal distribution.

Panel (l) - (o) are average charge-state distributions of carbon ($\mathrm{\langle Q\rangle C}$), oxygen ($\mathrm{\langle Q\rangle O}$), magnesium ($\mathrm{\langle Q\rangle Mg} $), and silicon ($\mathrm{\langle Q\rangle Si} $), respectively. Combined with $\mathrm{\langle Q\rangle Fe}$, and taking all elements as a whole, MCs have the highest charge-state, followed by SIMFR-SUN, and finally SIMFR-SW. The differences among MCs, SIMFR-SUN, and SIMFR-SW by ionic species are as follows: $\mathrm{\langle Q\rangle Fe}$ $>$ $\mathrm{\langle Q\rangle Si}$ $>$ $\mathrm{\langle Q \rangle Mg}$ $>$ $\mathrm{\langle Q\rangle O}$ $>$ $\mathrm{\langle Q\rangle C}$, which are consistent with the rank of elements atomic number from large to small.

Figure \ref{fig:abundance} is elemental abundances of specific ions relative to O. Panel (p) - (s) are He, Ne, C, and Mg sequentially. They are organized by First ionization potential \citep[FIP]{1995Sci...268.1033G}. As shown in the Panel(s), SIMFR-SUN are closer to MCs in low-FIP element (FIP $\leq 10$ eV, e.g. Mg, Si, Fe) abundances (Si/O and Fe/O are not shown), and they are slightly higher than SIMFR-SW, which are closer to those in the SW (referring to \citet{2020ApJ...893..136H} for SW). For high-FIP elements, compared to MCs, both populations of SIMFRs exhibit abundance depletion in Ne/O, similar He/O, and enhancement in C/O.

To quantify the similarities between MC and the two populations of SIMFR, we used two performance metrics: (1) the Euclidean distance (the sum of the pairwise distance between each value of MC and SIMFR in the distribution series) and (2) the variance. Results show that there are shorter Euclidean distances and smaller variance differences between MC and SIMFR-SUN compared with those between MC and SIMFR-SW in all of the parameters except $\left|B\right|$ and C/O (Table 1).

In summary, for most of the parameters in this study, especially $V_{rad}$, $N_{p}$, ion charge-states, and low-FIP element abundances, two populations of SIMFRs have marked differences and the SIMFR-SUN tend to be MC-like.

\section{Discussion and Conclusions } \label{sec:conclusion}

In this paper, following the controversy over whether two populations of SIMFRs exist in the SW, we presented a first comprehensive, long-term, statistical survey of the plasma and composition distribution inside SIMFRs, and made a comparison with MCs. The SIMFRs were divided into two categories according to SIMFR was observed far away from (close to) the HCSs and containing (without) the CSEs. This classification is expected to obtain two types of SIMFRs originating from the Sun and the SW, respectively. Although our study is based on a simple cylindrical flux-rope model, the distribution trend of internal parameters can be extracted. On the other hand, it is difficult to construct an average profile if the model is too complicated.

Results indicate that SIMFR-SUN and SIMFR-SW have noticeable differences. Compared to SIMFR-SW, SIMFR-SUN show faster and more fluctuant $V_{rad}$, lower $N_{p}$ in the center, higher ion charge-states, including $\mathrm{\langle Q\rangle Fe}$ and some charge-state ratios which are sensitive to heating near the sun, and higher low-FIP element abundances. These internal characteristics of SIMFR-SUN are closer to MCs than SIMFR-SW.
%

The formation locations partly determine the interplanetary properties of flux-ropes. It is well known that the MCs are a subset of ICMEs which originated from the Sun, and that the SIMFR-SUN are also expected to form in the Sun. Therefore, it is not surprising that SIMFR-SUN and MCs have some similar properties such as the ion charge-states. Concretely speaking, because $\mathrm{\langle Q\rangle Fe}$, $\mathrm{O^{7+}/O^{6+}}$, $\mathrm{Fe^{\geq16+}/Fe_{total}}$,  $\mathrm{C^{6+}/C^{4+}}$, and $\mathrm{C^{6+}/C^{5+}}$ are continuously affected by heating processes until become frozen-in then never change beyond a few solar radii in SW \citep[e.g.][]{1999JGR...10417005K,2003ApJ...582..467C}, they are proper identifiers of the heating experience and the diagnostics of the low-corona temperature. SIMFR-SUN show obviously higher ion charge-states than SIMFR-SW, implying that the SIMFR-SUN have went through hotter heating experience than SIMFR-SW. In addition, it is noteworthy that for ion charge-states in SIMFR-SUN, the sunward side is higher than the earthward side. This phenomenon is also found inside MCs \citep{2020ApJ...893..136H}, which suggested that the flare heated materials are ejected with the CMEs. Considering that the enhancement of heavy ions charge-states, including $\mathrm{O^{7+}/O^{6+}}$  $\mathrm{Fe^{\geq16+}/Fe_{total}}$, $\mathrm{Fe^{\geq16+}/Fe_{total}}$,  $\mathrm{C^{6+}/C^{4+}}$ and $\mathrm{C^{6+}/C^{5+}}$ ratio, is typically associated with CMEs \citep{2001JGR...10610597H,2001JGR...10629231L}, it is very likely that SIMFR-SUN are originating from the Sun like MCs, and heated by fares during the process of eruption. In all of the properties studied here, $\left|B\right|$ shows SIMFR-SW that are more similar to the MCs than the SIMFR-SUN, which should be attributed to the fact that SIMFR-SUN were born further away from the Earth orbit than SIMFR-SW, while $\left|B\right|$ in the SIMFRs decreases with the heliocentric distance \citep{2020ApJ...894..120M}.


The interplanetary properties of SIMFRs are also associated with the propagation experience in the SW, which caused SIMFR-SUN and SIMFR-SW to show some common properties in our results. The two population SIMFRs have comparable $T_{p}$, He/P, element compositions, and no significant expand. Besides, compared to MCs, both of them show lower $\left|B\right|$, with higher $T_{p}$ and plasma $\beta$. These common properties may mostly be caused by the same physics applied to SIMFR embedded in a common SW environment. For example, there is no expansion for both SIMFRs because SIMFRs are easily affected by this local SW pressure and store less free energy due to their small scales. The low $T_{p}$ of MCs is caused by expansion, while for SIMFRs, no expansion, thus miss the low $T_{p}$ characteristic. The higher plasma $\beta$ (compared to MCs) is caused by the lower $\left|B\right|$ and the higher $T_{p}$. These parameters are very close to those in SW (refer to the SW status in \citet{2020ApJ...893..136H}), which should be caused by the constant interaction and assimilation by the SW. In summary, these common properties can be explained by the small-scale and the interaction with the surrounding SW. \citet{2020ApJ...894...25C} also suggested that most SIMFRs properties exhibit radial decays from the Sun near the ecliptic plane. Thus, to further explore the SIMFR-SUM properties, weaken propagation effects are necessary. Presently, NASA PSP and ESA-NASA Solar Orbiter are expected to provide SIMFRs which are closer to the Sun.

To sum up, SIMFR-SUN and SIMFR-SW have distinct differences in plasma parameters, particularly internal ion charge-states distribution. SIMFR-SUN tend to be MC-like. Although there are also some common properties between the two types of SIMFRs, these commonalities can be explained by the interaction with the surrounding SW. SIMFR-SUN and MCs should all originate from the corona, SIMFR-SUN and SIMFR-SW may have different sources. In other words, our results supplement evidence in favor of the view that there are at least two sources for SIMFRs.


\acknowledgments

This work is funded by the grants from the Strategic Priority Research Program of CAS with grant XDA-17040507, and the National Science Foundation of China (NSFC 11533009, 11973086). This work is also funded by the Project Supported by the Specialized Research Fund for Shandong Provincial Key Laboratory. In addition, we are also grateful to the One Belt and One Road Scientific Project of the West Light Foundation, CAS. All the ACE and WIND data are from NASA CDAWeb. We are grateful to the data provided by the NASA/GSFC. We thank an anonymous reviewer for the constructive comments that are helpful in improving our study.


\listofchanges
\end{document}